\PassOptionsToPackage{table,svgnames,dvipsnames}{xcolor}
\documentclass[sigconf]{acmart}

\usepackage{caption} 
\usepackage{multirow}
\usepackage{makecell}       
\usepackage{graphicx}       
\usepackage{xcolor}         
\usepackage{placeins}

\AtBeginDocument{%
  }

\setcopyright{acmlicensed}
\copyrightyear{2018}
\acmYear{2018}
\acmDOI{XXXXXXX.XXXXXXX}
\acmConference[MM '25]{Proceedings of the 33rd ACM International Conference on Multimedia}{October 27--31, 2025}{Dublin, Ireland}

\acmISBN{978-1-4503-XXXX-X/2018/06}

\acmSubmissionID{6522}

\begin{document}

\renewcommand\footnotetextcopyrightpermission[1]{}
\settopmatter{printacmref=false}

\title{PRM-BAS: Enhancing Multimodal Reasoning through PRM-guided Beam Annealing Search}

\author{Pengfei Hu}
\authornote{Equal contributions.}  
\author{Zhenrong Zhang}
\authornote{Work done during an internship at iFLYTEK Research.}  
\authornotemark[1]
\affiliation{%
  \institution{University of Science and \\Technology of China}
  \city{}
  \state{}
  \country{}
}

\author{Qikai Chang}
\author{Shuhang Liu}
\affiliation{%
	\institution{University of Science and \\Technology of China}
	\city{}
	\state{}
	\country{}
}

\author{Jiefeng Ma}
\author{Jun Du}
\authornote{Corresponding author.}
\affiliation{%
	\institution{University of Science and \\Technology of China}
	\city{}
	\state{}
	\country{}
}

\author{Jianshu Zhang}
\author{Quan Liu}
\affiliation{%
	\institution{iFLYTEK Research}
	\city{}
	\state{}
	\country{}
}

\author{Jianqing Gao}
\author{Feng Ma}
\affiliation{%
	\institution{iFLYTEK Research}
	\city{}
	\state{}
	\country{}
}

\author{Qingfeng Liu}
\affiliation{%
	\institution{University of Science and \\Technology of China}
	\city{}
	\state{}
	\country{}
}

\renewcommand{\shortauthors}{Hu ang Zhang et al.}

\begin{abstract}
Recent work increasingly focuses on improving the reasoning capabilities of Multimodal Large Language Models (MLLMs). Among existing methods, Process Reward Models (PRMs) stand out for offering dense, step-wise supervision to guide intermediate reasoning. However, how to effectively integrate PRMs into search strategies remains an open question.
In this paper, we introduce $\textbf{PRM-BAS}$ ($\textbf{PRM}$-Guided $\textbf{B}$eam $\textbf{A}$nnealing $\textbf{S}$earch), a lightweight approach for PRM-guided reasoning that dynamically adjusts beam size—starting with a broader search space and gradually narrowing it as contextual information accumulates, thereby balancing performance and efficiency.
We further propose a unified framework for data construction and PRM training. Specifically, we construct the PRM-BAS-300k dataset by selecting 300k questions from existing datasets and performing rollouts at each step to estimate the probability of reaching a correct final answer. The PRM is then trained using a combination of value loss for absolute action quality and rank loss for relative action quality.
Extensive experiments on challenging multimodal reasoning benchmarks demonstrate that PRM-BAS significantly improves reasoning performance while maintaining low computational cost. Moreover, it generalizes well across different model scales and architectures, showcasing strong robustness and plug-and-play capability.
\end{abstract}

	



\maketitle

\begin{figure}[htb]
	\centering
	\includegraphics[width=0.9\linewidth]{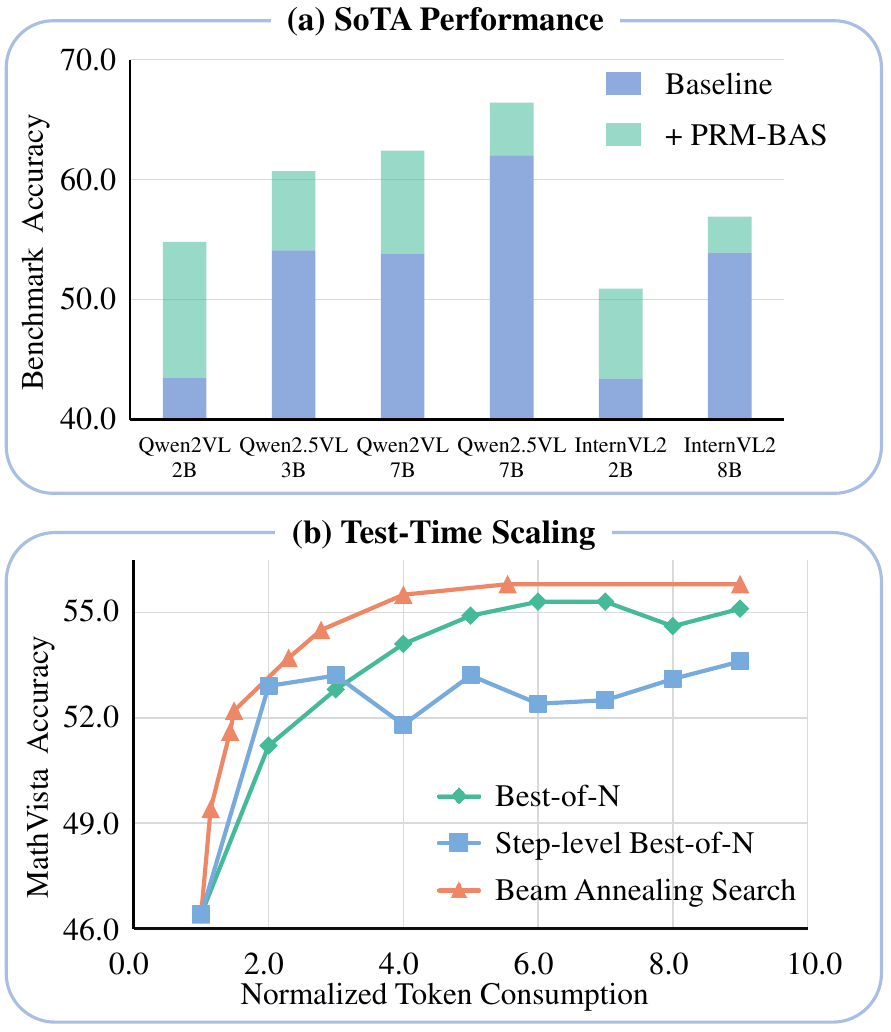}
	\caption{(a) Average accuracy across common benchmarks. (b) Test-Time Scaling curves under different token consumption ratios relative to single-shot inference.}
	\label{fig:teaser}
\end{figure}

\section{Introduction}
Large language models, such as OpenAI o1 \cite{o1} and DeepSeek-R1  \cite{ds-r1}, have shown strong abilities in in-depth reasoning. These models have achieved success in various NLP domains including math problem solving and code generation, proving the effectiveness of Test-Time Scaling (TTS) for language models. 

Inspired by the success of LLMs, the research community has recently turned to building Multimodal Large Language Models (MLLMs) that combine both language and vision to support deeper reasoning. Early efforts focus on designing tailored prompts to encourage the generation of chain-of-thought rationales that enhance reasoning capabilities \cite{cantor}. Subsequent works, such as Insight-V \cite{insight-v}, Virgo \cite{virgo}, and LLaVA-CoT \cite{llava-o1}, have made attempts to reach this goal by collecting high-quality long-chain reasoning data from stronger models \cite{gpt4o,ds-r1,qvq}, and then updating MLLMs through Supervised Fine-Tuning (SFT). Mulberry \cite{mulberry} further advances this approach by applying collective Monte Carlo Tree Search (MCTS) to explore reasoning paths that lead to correct answers. Recently, inspired by DeepSeek-R1's emergent abilities in complex reasoning \cite{ds-r1}, researchers have adopted R1-style reinforcement learning methods. Notable examples include Vision-R1 \cite{vision-r1} and Vision-RFT \cite{visual-rft}, which significantly improve reasoning performance. 

Beyond approaches that directly update model parameters, another line of work leverages search algorithms guided by reward models to enhance reasoning without modifying the base model \cite{liu2025inference}. These reward models generally fall into two categories: Outcome Reward Models (ORMs) and Process Reward Models (PRMs). ORMs assign an overall score to the final output and are typically used with Best-of-N (BoN) sampling, where the response with the highest reward is selected. However, due to their reliance on delayed feedback, ORMs struggle with credit assignment and evaluating the quality of intermediate reasoning steps. In contrast, PRMs offer step-level reward signals, which are highly valuable for tackling challenging reasoning problems \cite{lightman2023let,wang2023math}. Despite considerable efforts, three core challenges remain: \textbf{PRM-guided search}, \textbf{data construction}, and \textbf{PRM training}. For \textbf{PRM-guided search}, common strategies include BoN sampling \cite{visualprm,llava-o1}, Monte Carlo Tree Search (MCTS) \cite{rest-mcts,wan2024alphazero}, and beam search \cite{chen2024alphamath}. BoN only evaluates complete responses, limiting its ability to guide intermediate reasoning steps. MCTS and beam search, by contrast, offer stronger performance through step-wise exploration, but at the cost of high computational overhead, which restrict their scalability in real-world applications. For \textbf{data construction}, while PRMs can provide fine-grained feedback at each reasoning step, obtaining accurate step-by-step annotations remains difficult. Early work relies on human labeling \cite{lightman2023let}, which is costly and difficult to scale. Subsequent studies attempt automated annotation using advanced search algorithms like MCTS \cite{rest-mcts}, allowing LLMs or MLLMs to generate their own reasoning trees. However, 
the number of simulations per node varies, and only nodes with a high number of simulations can be used to provide reliable signals, making the process ineffective. Regarding \textbf{PRM training}, existing methods usually use binary labels to indicate whether each step is correct \cite{visualprm,qwenprm}. However, this can be ambiguous, especially as modern MLLMs increasingly depend on long chains of reasoning \cite{vision-r1,zhou2025r1}, where correct final answers may arise from incorrect intermediate steps through self-verification and reflection \cite{ds-r1}. Moreover, these methods often overlook the fact that although different actions can all lead to a correct final answer, their probability of success may vary significantly.

To address these challenges, we propose \textbf{PRM-BAS} (\textbf{PRM}-Guided \textbf{B}eam \textbf{A}nnealing \textbf{S}earch), an efficient framework to enhance the reasoning ability of MLLMs. For \textbf{PRM-guided search}, PRM-BAS adopts a dynamic beam size strategy, gradually reducing the beam size as reasoning progresses—unlike conventional beam search, which maintains a fixed beam size throughout. This design is based on the following insight: in the early steps, limited context makes it difficult for the PRM to reliably evaluate partial reasoning paths. As such, a larger beam size is initially required to provide the base model with sufficient exploration space and tolerance for suboptimal steps. As reasoning proceeds and more contextual information becomes available, the PRM gains a better understanding of the current state, allowing for a gradual reduction in beam size to reduce computational overhead. A detailed analysis of this motivation is discussed later. For \textbf{data construction}, we firstly sample approximately 300k question-answer pairs from the existing dataset \cite{shi2024math,m3cot,chartqa}, filtering out most multiple-choice and true-false questions to serve as the source for our training data. To improve sampling efficiency and ensure consistency with the PRM-BAS strategy, we directly perform rollouts at each reasoning step. Specifically, at each step, the policy model samples different action candidates, for each of which we perform $N$ full rollouts to complete the reasoning path. The candidate with the highest average success rate is selected for the next step. Regarding \textbf{PRM training}, our PRM directly employs the average success rate from rollouts as the training target. Additionally, we use a combination of value loss \cite{chen2024alphamath} and ranking loss \cite{wang2023math,luo2024improve}, which learn both the absolute quality of actions and their relative quality compared to alternatives.

We conducted experiments on several widely used and challenging datasets, covering domains from general and mathematical reasoning to visual illusion, and multidisciplinary understanding. As shown in Figure \ref{fig:teaser} (a), PRM-BAS significantly improves the reasoning performance of existing MLLMs on MathVista \cite{mathvista}, MathVision \cite{mathvision}, ChartQA \cite{chartqa} and M3CoT \cite{m3cot}. Furthermore, as shown in Figure \ref{fig:teaser} (b), we compared the proposed beam annealing search with BoN and step-level BoN under the TTS setting. Beam annealing search consistently achieves better reasoning accuracy than both baselines under comparable computational budgets. In addition, we validate the generalization ability of PRM-BAS across different model scales and architectures.

In summary, the main contributions of this work are as follows: \begin{itemize} 
	\item We propose beam annealing search, an efficient yet effective algorithm specifically designed for PRM-guided reasoning. 
	\item We further design a unified pipeline for data construction and PRM training, aligned with PRM-guided reasoning. 
	\item We validate the effectiveness of our approach through extensive experiments across multiple benchmarks.
\end{itemize}


\begin{figure*}[t]
	\centering
	\includegraphics[width=0.97\linewidth]{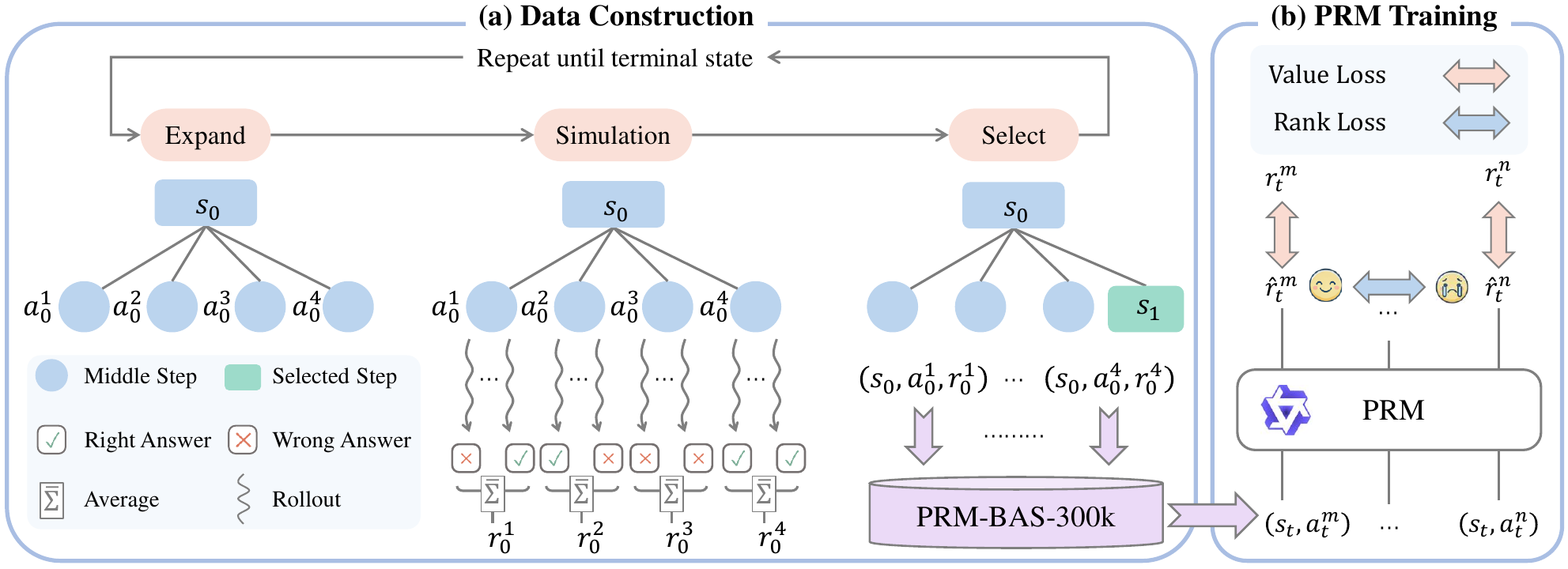}
	\caption{(a) The illustration of data construction. (b) The illustration of PRM training. }
	\label{fig:pipeline}
\end{figure*}

\section{Related Work}
To improve the reasoning abilities of MLLMs, existing research has explored three main directions: prompt-based, learning-based, and search-based methods. We review each category below.

\subsection{Prompt-based Methods}

Prompt-based methods are train-free approaches that design prompts to make MLLMs take on different roles, generating intermediate reasoning results or strategies in a workflow manner. Cantor \cite{cantor} assigns different tasks to a single MLLM using various expert identities and task instructions, exploring the potential of an MLLM to act as different experts. It breaks down the visual reasoning task into two steps: decision generation and execution. In the first step, the multimodal model is prompted to take on roles such as principle analysis, module selection, and task allocation. In the second step, the model generates corresponding high-level visual features based on task analysis. Finally, the results of the subtasks are synthesized and summarized to provide the final answer. CCot \cite{ccot} further utilizes scene graphs to formally represent the results of visual reasoning, offering a highly structured representation of visual objects, relationships, and attributes within an image. Astar \cite{astar} introduces six atomic reasoning actions, called "thought cards," which simulate human-like cognitive behaviors, such as problem decomposition and reasoning step reflection. After deriving reference reasoning patterns to construct multiple thought cards, Astar retrieves the card most similar to the target problem during inference and then performs visual reasoning.
Although these methods are relatively easy to implement, they require customized prompts, which limits their generalization ability. Additionally, the performance improvement is often constrained \cite{lin2025mind}.

\subsection{Learning-based Methods}
Learning-based methods typically begin by constructing a training dataset that includes reasoning chains, then applying SFT or reinforcement learning to optimize MLLMs. We introduce these two components separately below:

\textbf{Data Construction.} The goal at this stage is to collect data with reasoning chains for future learning. Based on the source of labels, these methods can be divided into two categories. The first category uses powerful teacher models \cite{gpt4o,ds-r1} to generate chain-of-thought outputs and answers \cite{llava-o1,zhang2024improve,shao2024visual,thawakar2025llamav}. Some also employ robust open-source models to filter low-quality data \cite{dong2024insight,wang2024enhancing}. To leverage the reasoning capabilities of existing LLMs, some methods convert images into captions, which are then input alongside the original questions into LLMs to generate solutions in a chain-of-thought format \cite{chen2025r1v,vision-r1,yang2025r1}. The second category of methods \cite{cheng2024vision} uses the base model itself to sample and generate reasoning paths, followed by iterative self-training. Mulberry \cite{mulberry} improves the diversity of reasoning paths by using MCTS with a policy ensemble of multiple MLLMs. It also constructs reflective reasoning paths that transition from incorrect to correct reasoning steps.

\textbf{Model Training.} After collecting data, either supervised fine-tuning (SFT) or reinforcement learning (RL) is applied to optimize model performance. Based on the training strategy, existing methods can be categorized into three types. The first type relies solely on SFT \cite{llava-o1,mulberry}, using reasoning paths as training targets. Some approaches incorporate curriculum learning \cite{thawakar2025llamav}, starting with simple tasks such as image captioning and progressing to more complex multimodal reasoning tasks. Iterative self-training is also adopted \cite{mulberry,cheng2024vision}, where the model is continuously fine-tuned on its own generated rationales. The second type combines SFT with RL. For example, some methods employ Direct Preference Optimization (DPO) to fine-tune policy models based on preference-labeled data \cite{zhang2024improve,wang2024enhancing}, while Insight-V \cite{insight-v} further performs multiple rounds of sampling and DPO to better simulate online reinforcement learning. Other methods define rule-based reward functions and apply Group Relative Policy Optimization (GRPO) to encourage more reliable and generalizable reasoning \cite{yang2025r1,vision-r1}. The third type omits SFT entirely and trains models using RL alone. Representative examples include R1-zero \cite{zhou2025r1} and Visual-RFT \cite{visual-rft}, which directly optimize reasoning capabilities through reinforcement learning.

\subsection{Search-based Methods}
Search-based methods aim to improve the reasoning ability of MLLMs by progressively selecting better actions from candidate options during inference, thus enhancing the overall reasoning performance step by step. The effectiveness of such methods largely depends on the quality of the reward model used to guide the search toward correct answers. Existing reward models can be broadly categorized into ORMs and PRMs. ORMs evaluate the quality of final outputs \cite{wang2024interpretable,zhang2024generative}, but suffer from delayed feedback and credit assignment issues, making it difficult to identify which specific reasoning steps contributed to the final outcome. In contrast, PRMs assess each intermediate step, offering denser reward signals throughout the reasoning process \cite{uesato2022solving,lightman2023let}. Prior studies have shown that PRMs outperform ORMs in guiding reasoning \cite{lightman2023let,wang2023math}, making PRM-guided search the primary focus in recent research on MLLMs. However, using PRMs to guide MLLMs remains challenging. Common search strategies include BoN, step-level BoN, MCTS, and beam search. BoN only evaluates completed responses, offering reward signals too late to influence intermediate steps. Step-level BoN improves upon this by selecting the best candidate at each step, while it can be unreliable in early stages, leading to suboptimal trajectories. MCTS and beam search offer better performance via stepwise guidance, but are often computationally intensive and slow, limiting their practicality. In this paper, we explore a search strategy that is both effective and efficient. Alongside this, we present a comprehensive pipeline that covers data construction and PRM training, to facilitate more robust step-by-step reasoning in MLLMs.

\begin{figure*}[t]
	\centering
	\includegraphics[width=0.97\linewidth]{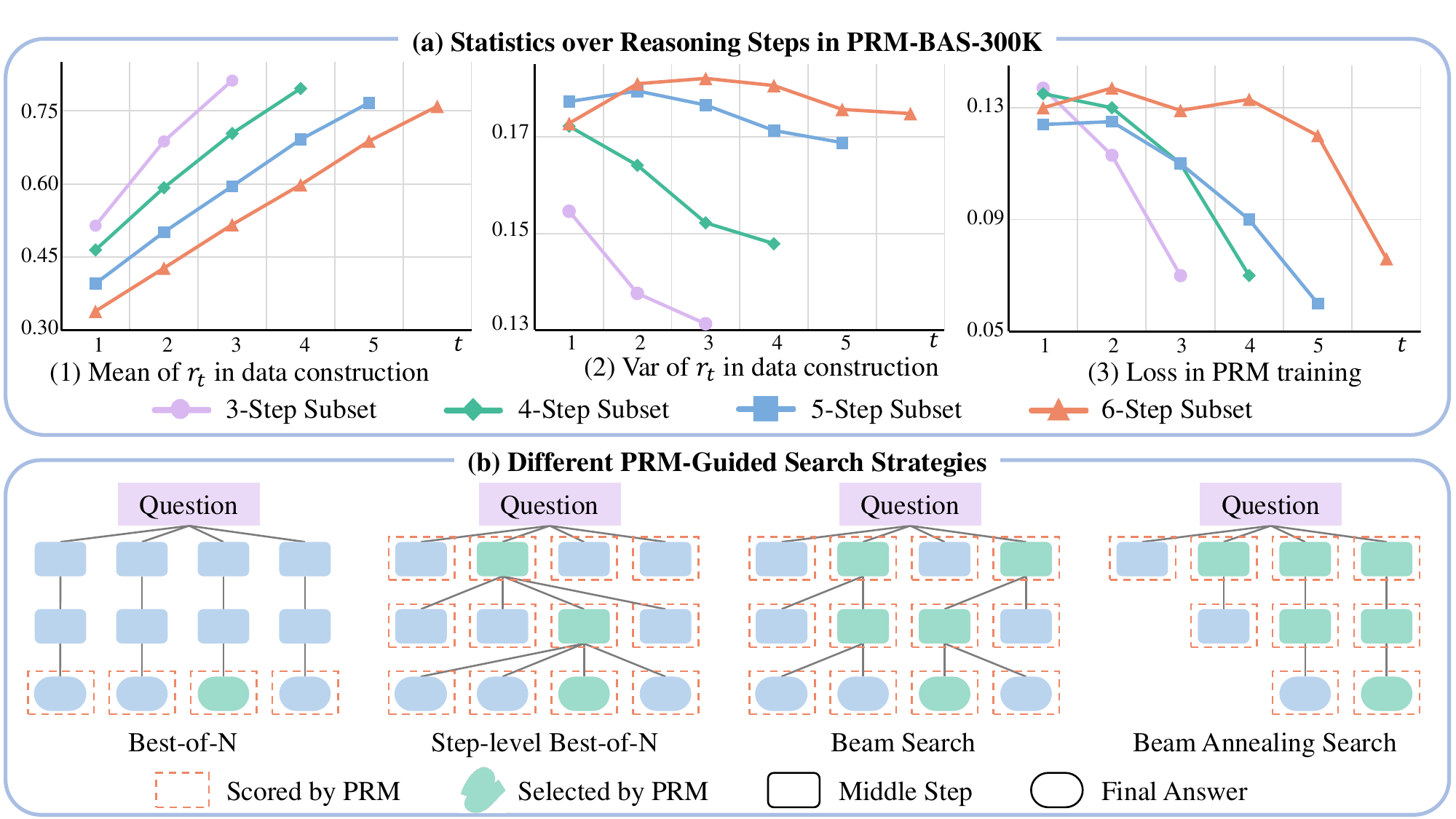}
	\caption{(a) The statistics of step-wise reward during data construction, including the mean and variance of $r_t$, and the training loss of PRM across reasoning steps. (b) The comparison of different PRM-guided search strategies. }
	\label{fig:bas}
\end{figure*}

\section{Method}
\subsection{Problem Formulation} 
We consider an MLLM, parameterized by $\theta$, denoted as a policy $\pi_\theta$ and modeled as a conditional probability distribution $p_\theta$. The model takes a multimodal input consisting of an image $\boldsymbol{I}$ and a question $\boldsymbol{x}$, and generates an answer sequence $\boldsymbol{y}$. We then formulate the process of PRM-guided reasoning as a Markov Decision Process (MDP). Prior works typically define actions at either the token or sentence level \cite{rest-mcts,qwenprm,wan2024alphazero}. Token-level actions, which treat each token as an atomic decision, are too fine-grained to offer meaningful learning signals. Sentence-level actions, such as splitting by the delimiter ``\texttt{\textbackslash n\textbackslash n}'', align better with human reasoning but often generalize poorly and introduce inefficiencies in real-world applications. Instead, we simply define each action as a fixed-length segment of $\mathcal{L}$ tokens (set to $\mathcal{L} = 30$ in our experiments), resulting in the answer sequence being represented as $\boldsymbol{y} = [\boldsymbol{a}_0, \boldsymbol{a}_1, \dots, \boldsymbol{a}_{T-1}]$, where $|\boldsymbol{a}_t| = \mathcal{L}$ and $0 \leq t \leq T-1$. The MLLM modeling the conditional distribution in an autoregressive manner:
\begin{equation}
	p_\theta(\boldsymbol{y} \mid \boldsymbol{I}, \boldsymbol{x}) = \prod_{t=0}^{T-1} p_\theta(\boldsymbol{a}_t \mid \boldsymbol{I}, \boldsymbol{x}, \boldsymbol{y}_{<t})
\end{equation}
 Under this formulation, the MDP is defined as $(\mathcal{S}, \mathcal{A}, \mathcal{T}, \mathcal{R}, \gamma)$:
\begin{itemize}
	\item \textbf{State} $\boldsymbol{s}_t \in \mathcal{S}$: The state at time step $t$ includes the image $\boldsymbol{I}$, the question $\boldsymbol{x}$, and the partial answer $[\boldsymbol{a}_0, \dots, \boldsymbol{a}_{t-1}]$. The initial state $\boldsymbol{s}_0$ corresponds to the input $\boldsymbol{I}$ and $\boldsymbol{x}$.
	\item \textbf{Action} $\boldsymbol{a}_t \in \mathcal{A}$: An action corresponds to generating a segment of $\mathcal{L}$ tokens as defined above.
	\item \textbf{Transition} $\mathcal{T}(\boldsymbol{s}_t, \boldsymbol{a}_t) \rightarrow \boldsymbol{s}_{t+1}$: The next state $\boldsymbol{s}_{t+1}$ is obtained by appending $\boldsymbol{a}_t$ to the current sequence.
	\item \textbf{Reward} $r_t = \mathcal{R}(\boldsymbol{s}_t, \boldsymbol{a}_t)$: A scalar score assessing the immediate reward of action $\boldsymbol{a}_t$ in the context of state $\boldsymbol{s}_t$.
\end{itemize}
$\gamma$ is omitted because it is not relevant in our setting. Our PRM is designed to estimate the probability of eventually reaching a correct answer rather than evaluating the correctness of each step. This design follows the trend in modern MLLMs, which increasingly rely on long reasoning chains where incorrect steps may still lead to correct answers through reflection and verification.

\subsection{Data Construction}

The effectiveness of PRM largely relies on the quality of its training data. However, manually annotating accurate step-level supervision is both costly and difficult to scale~\cite{lightman2023let}. Therefore, we propose an automated step-level rollout-based sampling strategy to construct our dataset, PRM-BAS-300k, as illustrated in Figure~\ref{fig:pipeline} (a).

\textbf{Source Dataset Collection.} We collect question-answer pairs from the MathV360K \cite{shi2024math}, which spans a wide range of tasks, including free-form question answering, geometry problem solving, math word problems, textbook QA, and visual QA. To ensure coverage of diverse reasoning scenarios, we further incorporate M3CoT \cite{m3cot} to increase multi-step chain-of-thought samples, as well as chart data from ChartQA \cite{chartqa}. To improve annotation reliability, we exclude most multiple-choice and true/false questions, which often lead to inconsistencies. For example, when the model generates an answer that is not among the provided options, it may still guess an option anyway, possibly matching the target by chance. We retain only a small portion of such questions to preserve diversity. The final number of selected question-answer pairs is approximately 300k.

\textbf{Rollout-based Step Sampling. \label{sec:data}}  
Previous approaches often rely on complex search methods like MCTS, which can only provide reliable signals at well-explored nodes, limiting efficiency. To simplify this, we use a step-level sampling strategy. Given a state $\boldsymbol{s}_t$, we sample $M$ candidate actions $\boldsymbol{a}_t^1, \dots, \boldsymbol{a}_t^M$. For each candidate $\boldsymbol{a}_t^i$, we perform $N$ rollouts,  yielding $N$ final answers. Each final answer is compared to the ground truth and assigned a score of 1 (correct) or 0 (incorrect). The average score over $N$ rollouts, denoted $ {r}_t^i $, is used as the target. The resulting triplet $(\boldsymbol{s}_t, \boldsymbol{a}_t^i, r_t^i)$ is added to the training set $\mathcal{D}$. We then select the candidate with the highest ${r}^i_t$ to transition to the next state $\boldsymbol{s}_{t+1}$, and repeat this process until an end-of-sequence token is generated.

\textbf{Efficient Sampling Adjustment.} 
Before scaling up to full dataset construction, we first apply the above sampling strategy to a randomly selected subset of 5k question-answer pairs, denoted as $\mathcal{D}_{\text{tiny}}$. To analyze reward dynamics at different reasoning stages, we group samples based on the number of reasoning steps in the completed response $\boldsymbol{y}$. Specifically, we define an $n$-step subset to include all samples where $(n-1)\mathcal{L} \leq |\boldsymbol{y}| < n\mathcal{L}$, with $|\boldsymbol{y}|$ indicating the total number of tokens in the completed response and $\mathcal{L} = 30$ representing the fixed token length per action. For clarity, we focus on subsets with $n = 3, 4, 5, 6$, and visualize the step-wise evolution of reward statistics in Figure~\ref{fig:bas} (a.1/2). Two key patterns emerge from this analysis: (1) The average reward $r_t$ tends to increase with step index $t$, suggesting that later actions are more likely to reach correct final answers, thus confirming the validity of our sampling process. (2) The variance of $r_t$ is higher at earlier steps, indicating greater uncertainty and a larger search space during early reasoning, while later steps become more stable and deterministic. These observations motivate a dynamic sampling scheme across reasoning steps. For earlier steps, we use larger values of $M$ and $N$ to ensure more reliable reward estimation. For later steps, we reduce $M$ and $N$ to improve sampling efficiency.
Additionally, for $\boldsymbol{s}_t$ with action candidates $a_t^1,...a_t^M$, we control the number of positive actions (${r_t^i} > 0.5$) and negative ones (${r_t^i} \leq 0.5$) such that the ratio of the more frequent type to the less frequent type does not exceed 3:1. If all actions belong to one type, we randomly retain at most 3 actions. 
By applying the above adjustment to the full question-answer pairs, we finally obtain the dataset, PRM-BAS-300k.


\subsection{PRM Training \label{sec:train}}  
Inspired by the strong reasoning capabilities of long CoT demonstrated in LLMs~\cite{ds-r1,o1}, recent MLLMs have adopted similar strategies to improve multimodal reasoning performance~\cite{vision-r1,zhou2025r1}, which introduces a new challenge: models can sometimes arrive at correct final answers based on incorrect intermediate reasoning steps through reflection and inspection. As a result, unlike previous approaches that directly learn the binarized correctness~\cite{qwenprm}, we train the PRM to estimate the likelihood that the current state will lead to a correct final outcome, as shown in Figure \ref{fig:pipeline} (b).

Specifically, given a state $\boldsymbol{s}_t$ and a set of $M$ candidate actions $\boldsymbol{a}^1_t, \dots, \boldsymbol{a}^M_t$, the PRM $q_\phi$ is trained to predict the reward for taking the $i$-th action $\boldsymbol{a}^i_t$ in the context of state $\boldsymbol{s}_t$, where $1 \leq i \leq M$. $q_\phi$ is initialized from the base model, with the language modeling head replaced by a reward head—an MLP layer that outputs a scalar for each token. The scalar prediction at the last token, denoted $\hat{r}^i_t$, is used as the estimated reward. To optimize the PRM, we first apply a binary cross-entropy loss to learn the absolute value of $i$-th action based on the corresponding ground-truth soft reward $r^i_t \in [0,1]$:
\begin{equation} \label{equ:value}
	\mathscr{L}_{\rm value} = -\frac{1}{T} \sum_{t=0}^{T-1} \frac{1}{M} \sum_{i=1}^{M} \big[ r^i_t \log \hat{r}^i_t + (1 - r^i_t) \log (1 - \hat{r}^i_t) \big]
\end{equation}

Additionally, we incorporate an auxiliary ranking loss to model the relative ordering among different action candidates, which has been validated as effective in other domains~\cite{lin2024understanding,sheng2023joint,yan2022scale}.
\begin{equation} \label{equ:rank}
	\mathscr{L}_{\rm rank} = -\frac{1}{T} \sum_{t=0}^{T-1}\frac{1}{|\mathcal{S}_t|}\sum_{(m, n) \in \mathcal{S}_t} \log \sigma(\hat{r}^m_t - \hat{r}^n_t)
\end{equation}
where the set $\mathcal{S}_t$ contains all index pairs $(m, n)$ such that $r^m_t - r^n_t > \delta$. We set $\delta = 0.3$ to suppress the influence of noisy comparisons. The overall loss function is defined with a weight $\lambda$ as:
\begin{equation}
	\mathscr{L} = \mathscr{L}_{\rm value} + \lambda \mathscr{L}_{\rm rank}
\end{equation}

\subsection{PRM-guided Search}
Balancing performance and efficiency in PRM-guided reasoning remains an open question. We identify two key characteristics of PRM-guided search. First, as illustrated in Figure~\ref{fig:bas} (a.2), \textbf{the variance of reward scores $r_t$ is high in early steps}, suggesting that the policy model faces a larger exploration space at the beginning of the reasoning process. Second, we visualize the training loss after one epoch at each step in Figure~\ref{fig:bas} (a.3), showing that \textbf{the loss is high in early steps}, indicating that the PRM struggles to accurately evaluate states when contextual information is limited. Motivated by these findings, we propose a new inference strategy: Beam Annealing Search (BAS). In early steps, we adopt a larger beam size to provide redundancy and improve tolerance to PRM estimation errors. As reasoning progresses, the beam size is gradually reduced to enhance computational efficiency. Assuming the beam size at the initial state $\boldsymbol{s}_0$ is $b_0$, the beam size at step $t$, denoted $b_t$, is updated according to the following annealing schedule: \begin{equation} b_t = \max(b_0 - kt, \epsilon) \end{equation} Here, $k$ is a hyperparameter controlling the annealing rate, and $\epsilon$ is a small positive constant that ensures sufficient diversity in later stages. Additionally, the commonly used hyperparameter in beam search, the expansion number, is typically set to 1 and thus omitted from the formula for clarity.

\begin{table*}
	\caption{Performance comparison across diverse multimodal reasoning benchmarks.}
	\label{tab:sota}
	\resizebox{0.97\linewidth}{!}{%
		
		\begin{tabular}{lccccccccc}
			\toprule
			\textbf{Method} & \textbf{ \makecell{Math\\Vista}}  & \textbf{\makecell{Math\\Vision}} & \textbf{\makecell{MathVerse\\VO}} & \textbf{\makecell{Dyna\\Math}} & \textbf{M3CoT} & \textbf{\makecell{Chart\\QA}} & \textbf{\makecell{Logic\\Vista}} & \textbf{\makecell{Science\\QA}} & \textbf{AVG.} \\
			\midrule
			\multicolumn{10}{c}{\textit{Closed-Source Model}} \\
			GPT-4o\cite{gpt4o} & 63.8 & 30.4 & 40.6 & 63.7 & 64.3 & 85.7 & 52.8 & - &  \\
			Claude-3.5 Sonnet\cite{anthropic2024claude} & 67.7 & 35.6 & 46.3 & 64.8 & - & 90.8 & 60.4 & - &  \\
			Gemini-2.0-Flash\cite{team2024gemini} & 70.4 & 43.6 & 47.8 & - & - & - & 52.3 & - &  \\
			\midrule
			\multicolumn{10}{c}{\textit{Open-Source Model}} \\
			DeepSeek-VL-7B\cite{lu2024deepseek} & 36.1 & - & - & 21.5 & - & 59.1 & - & - &  \\
			DeepSeek-VL2-MOE-4.5B\cite{wu2024deepseek} & 62.8 & - & - & - & - & 86.0 & - & - &  \\
			InternVL2-8B\cite{chen2024far} & 58.3 & 18.4 & 20.4 & 39.7 & 59.3 & 83.3 & 33.6 & 88.4 &  \\
			InternVL2.5-8B\cite{chen2024expanding} & 64.4 & 19.7 & 39.5 & - & - & 84.8 & - & - &  \\
			MiniCPM-Llama-V-2.5-8B\cite{yao2024minicpm} & 54.3 & 18.4 & 18.3 & - & - & - & 27.5 & - &  \\
			MiniCPM-V-2.6-8B\cite{yao2024minicpm} & 60.6 & - & 24.1 & - & 56.0 & - & - & 90.9 &  \\
			LLaVA-NeXT-8B\cite{liu2024llavanext} & 37.5 & - & - & 22.7 & - & 69.5 & - & - &  \\
			\midrule
			\multicolumn{10}{c}{\textit{Reasoning Model}} \\
			LLaVA-CoT-11B\cite{llava-o1} & 54.8 & - & - & - & - & - & - & - &  \\
			Mulberry-7B\cite{yao2024mulberry} & 63.1 & - & - & 45.1 & - & 83.9 & - & - &  \\
			\midrule
			Qwen2-VL-7B\cite{qwen2vl} & 58.2 & 16.3 & 30.8 & 48.3 & 57.8 & 83.0 & 35.0 & 80.1 & 51.2 \\
			\rowcolor{green!7}
			+ PRM-BAS & 67.2\raisebox{0.8ex}{\scriptsize\textcolor{green!60!black}{$\uparrow$ 9.0}} & 23.4\raisebox{0.8ex}{\scriptsize\textcolor{green!60!black}{$\uparrow$ 7.1}} & 41.3\raisebox{0.8ex}{\scriptsize\textcolor{green!60!black}{$\uparrow$ 10.5}} & 53.7\raisebox{0.8ex}{\scriptsize\textcolor{green!60!black}{$\uparrow$ 5.4}} & 72.3\raisebox{0.8ex}{\scriptsize\textcolor{green!60!black}{$\uparrow$ 14.5}} & 86.7\raisebox{0.8ex}{\scriptsize\textcolor{green!60!black}{$\uparrow$ 3.7}} & 41.0\raisebox{0.8ex}{\scriptsize\textcolor{green!60!black}{$\uparrow$ 6.0}} & 91.1\raisebox{0.8ex}{\scriptsize\textcolor{green!60!black}{$\uparrow$ 11.0}} & 59.6\raisebox{0.8ex}{\scriptsize\textcolor{green!60!black}{$\uparrow$ 8.4}} \\		
			\midrule
			Qwen2.5-VL-7B\cite{qwen2.5vl} & 68.2 & 25.1 & 46.3 & 57.1 & 67.6 & 87.2 & 43.8 & 81.6 & 59.6 \\
			\rowcolor{green!7}
			+ PRM-BAS & 72.9\raisebox{0.8ex}{\scriptsize\textcolor{green!60!black}{$\uparrow$ 4.7}} & 28.3\raisebox{0.8ex}{\scriptsize\textcolor{green!60!black}{$\uparrow$ 3.2}} & 51.5\raisebox{0.8ex}{\scriptsize\textcolor{green!60!black}{$\uparrow$ 5.2}} & 61.3\raisebox{0.8ex}{\scriptsize\textcolor{green!60!black}{$\uparrow$ 4.2}} & 75.2\raisebox{0.8ex}{\scriptsize\textcolor{green!60!black}{$\uparrow$ 7.6}} & 89.2\raisebox{0.8ex}{\scriptsize\textcolor{green!60!black}{$\uparrow$ 2.0}} & 45.5\raisebox{0.8ex}{\scriptsize\textcolor{green!60!black}{$\uparrow$ 1.7}} & 89.5\raisebox{0.8ex}{\scriptsize\textcolor{green!60!black}{$\uparrow$ 8.0}} & 64.2\raisebox{0.8ex}{\scriptsize\textcolor{green!60!black}{$\uparrow$ 4.6}} \\				
			\bottomrule
		\end{tabular}
		
	}
\end{table*}

\section{Experiments}
\subsection{Datasets}
We evaluate PRM-BAS on diverse benchmarks for multimodal reasoning. MathVista~\cite{mathvista}, MathVision~\cite{mathvision}, MathVerse~\cite{zhang2024mathverse}, DynaMath~\cite{dynamath}, and M3CoT~\cite{m3cot} focus on mathematical visual reasoning, requiring interpretation and inference over diagrams, charts, and multimodal content. ChartQA~\cite{chartqa} targets reasoning over structured visual data, such as bar and line charts. LogicVista~\cite{xiao2024logicvista} and ScienceQA~\cite{saikh2022scienceqa} evaluate logical inference and scientific knowledge understanding. For MathVista, we use \texttt{testmini} since the full test set labels are not publicly available. For DynaMath, we use \texttt{variant 1} to reduce computational cost.





\subsection{Implementation Details}

\textbf{Data Construction \& PRM Training.} We employ Qwen2-VL-7B~\cite{qwen2vl}, a relatively older model, as the base model for our PRM. This choice is made to ensure that performance improvements come from the effectiveness of PRM-BAS itself, rather than from a stronger reward model such as Qwen2.5-VL-7B \cite{qwen2.5vl}. The loss weight of rank loss $\lambda$ is set to 0.1. To reduce memory consumption and accelerate training, we adopt distributed training with mixed precision and gradient accumulation. PRMs are fine-tuned for 2 epochs on 32 Tesla A800 80GB GPUs with a global batch size of 1,024. We use AdamW~\cite{adamw} with a fixed learning rate of $5 \times 10^{-6}$. We adopt ZeRO~\cite{zeromemory} for memory-efficient full-parameter tuning.

\textbf{Beam Annealing Search.} By default, we set the initial beam size $b_0 = 12$, the annealing rate $k = 1$, and the minimum beam size $\epsilon = 2$ to provide sufficient exploration while maintaining efficiency. This setting yields token consumption roughly equivalent to BoN with $N = 8$. However, in cases where a fair comparison with other inference strategies is needed, we adjust these parameters to match overall token usage. 

In addition, to ensure fair and consistent evaluation, we adopt a unified prompt template for all experiments: \texttt{Please answer the question and provide the correct answer, e.g., 1, 2, 3, 4, at the end. Give step by step reasoning before you answer, and when you're ready to answer, please use the format "Final answer: ..."}.

\subsection{Comparison With State-of-the-Art Methods \label{sec:sota}}

To evaluate the effectiveness of PRM-BAS, we conduct comprehensive experiments using two strong baseline models, Qwen2-VL-7B and Qwen2.5-VL-7B, and compare them against a variety of recent MLLMs, as shown in Table~\ref{tab:sota}. We first observe that both Qwen2-VL-7B and Qwen2.5-VL-7B achieve substantial improvements when guided by PRM-BAS, demonstrating that our method can significantly enhance the performance of policy models. The token consumption introduced by PRM-BAS amounts to $7.2 \times $ to $8.7 \times$ that of the baseline across the eight evaluated datasets. Additionally, we compare PRM-BAS with both open-source and closed-source state-of-the-art models. Despite relying on only a 7B policy model, PRM-BAS outperforms most open-source MLLMs and achieves competitive results compared to some closed-source models.

We further compare our BAS with BoN, step-level BoN on MathVista excluding multiple-choice or true/false questions, and visualize the TTS results in Figure \ref{fig:teaser}. The x-axis represents the relative token consumption compared to single-shot inference. All methods use the same Qwen2VL-7B as the policy. For BAS, We vary $b_0$ from $1,2,...,14$ and $\epsilon$ from ${1,2}$ with a fixed annealing rate $k=1$. As the token budget increases, BAS continues to provide stable and incremental performance gains, and consistently outperforms both BoN and step-level BoN.

\begin{table}
	\caption{Ablation study configurations.}
	\label{tab:ablation_config}
	\renewcommand{\arraystretch}{1.1}
	\resizebox{0.9\linewidth}{!}{%
		\centering
		\begin{tabular}{l|cc|cc|cc}
			\toprule
			\multicolumn{1}{c}{} 
			& \multicolumn{2}{|c|}{\textbf{Training Data}} 
			& \multicolumn{2}{c|}{\textbf{Training Loss}} 
			& \multicolumn{2}{c}{\textbf{Labels}} \\
			
			\cmidrule(lr){2-3}
			\cmidrule(lr){4-5}
			\cmidrule(lr){6-7}
			
			\multicolumn{1}{c|}{} 
			& Outcome & Process 
			& Value & Rank 
			& Hard & Soft \\
			
			\midrule
			T1 & \checkmark &  & \checkmark & \checkmark &  & \checkmark \\
			T2 & \checkmark & \checkmark & \checkmark &  &  & \checkmark \\
			T3 & \checkmark & \checkmark & \checkmark & \checkmark & \checkmark & \\
			T4 & \checkmark & \checkmark & \checkmark & \checkmark &  & \checkmark \\
			\bottomrule
		\end{tabular}
	}
\end{table}

\begin{table*}
	\caption{Results of ablation study across multiple benchmarks and inference strategies.}
	\label{tab:ablation_result}
	\resizebox{0.99\linewidth}{!}{%
		\renewcommand{\arraystretch}{1.1}
		
		\begin{tabular}{l|ccccc|ccccc|ccccc}
			\toprule
			
			\multicolumn{1}{c}{} 
			& \multicolumn{5}{|c|}{\textbf{Best-of-N (N = 8)}} 
			& \multicolumn{5}{c|}{\textbf{Step-level Best-of-N (N = 8)}}
			& \multicolumn{5}{c}{\textbf{Beam Annealing Search}} \\

			\cmidrule(lr){2-6}
			\cmidrule(lr){7-11}
			\cmidrule(lr){12-16}

			\multicolumn{1}{c|}{} 
			& \makecell{Math\\Vista} & \makecell{Math\\Vision} & \makecell{Chart\\QA} & M3CoT & Avg. 
			& \makecell{Math\\Vista} & \makecell{Math\\Vision} & \makecell{Chart\\QA} & M3CoT & Avg.
			& \makecell{Math\\Vista} & \makecell{Math\\Vision} & \makecell{Chart\\QA} & M3CoT & Avg. \\
			
			\midrule
			T1 & 65.5 & 22.4 & 84.8 & 66.4 & 59.8 & 59.1 & 19.1 & 80.4 & 61.6 & 55.1 & 63.2 & 19.6 & 83.9 & 65.8 & 58.1 \\
			T2 & 65.5 & 22.8 & 85.7 & 69.3 & 60.8 & 68.2 & 21.2 & 84.9 & 68.9 & 60.8 & 65.9 & 21.8 & 85.0 & 71.1 & 61.0 \\
			T3 & 65.4 & 22.3 & 85.4 & 68.1 & 60.3 & 66.0 & 19.6 & 86.3 & 69.2 & 60.3 & 67.7 & 21.4 & 85.8 & 68.5 & 60.9 \\
			T4 & 67.4 & 22.8 & 84.2 & 71.4 & 61.5 & 67.5 & 20.6 & 89.0 & 67.7 & 61.2 & 67.2 & 23.4 & 86.7 & 72.3 & 62.4 \\
			\bottomrule
		\end{tabular}	
		
	}
\end{table*}

\subsection{Ablation Study}
By selectively removing modules, we construct four system variants: T1, T2, T3, and T4, as summarized in Table~\ref{tab:ablation_config}. Outcome supervision refers to training data that only includes the final answer, while process supervision involves data from intermediate reasoning steps. The value/rank loss is defined in Equations~\ref{equ:value}/\ref{equ:rank}, respectively. A soft label represents the estimated probability that a given action leads to a correct final result, whereas a hard label binarizes this probability to 0 or 1 based on a predefined threshold \cite{visualprm,qwenprm}. We evaluate all variants on MathVista, MathVision, ChartQA and M3CoT using three search strategies: BoN, step-level BoN, and the proposed BAS, with $N = 8$ for BoN to ensure comparable computational cost.

\textbf{Impact of Process Supervision.}  
We remove all process supervision from the training set, retaining only outcome supervision.  The resulting PRM is referred to as T1 (more precisely, T1 functions as an ORM). Compared to the fully supervised model T4, T1 shows a noticeable performance drop under the BAS and step-level BoN, confirming the critical role of process supervision. Interestingly, T1 also underperforms T4 under the BoN strategy, indicating that process supervision benefits both PRMs and ORMs.

\textbf{Impact of Rank Loss.}  
By setting $\lambda = 0$, we obtain T2, which relies solely on the value loss. T2 shows a consistent performance drop across all three search strategies compared to T4, suggesting that rank loss effectively helps the PRM distinguish the relative quality of different actions under the same state, which makes it a valuable complement to the value loss.

\textbf{Impact of Soft/Hard Labels.} Different from our PRM using soft labels (the estimated likelihood that an action leads to a correct final answer) as training targets, an alternative approach used in prior work \cite{wang2024math,dong2024rlhf} is to apply hard labels, where correctness is binarized using a threshold (typically 0~\cite{visualprm,qwenprm}). We compare these two strategies through models T3 (hard label) and T4 (soft label). Clearly, soft labels lead to better performance. The Qwen team~\cite{qwenprm} has shown that the performance of hard labels can be improved by data filtering such as LLM-as-a-judge. However, we choose not to adopt this approach, as modern MLLMs tend to generate long chains of reasoning in which incorrect intermediate steps may still lead to correct final answers through self-verification and reflection.

\begin{figure*}[htb]
	\centering
	\includegraphics[width=0.9\linewidth]{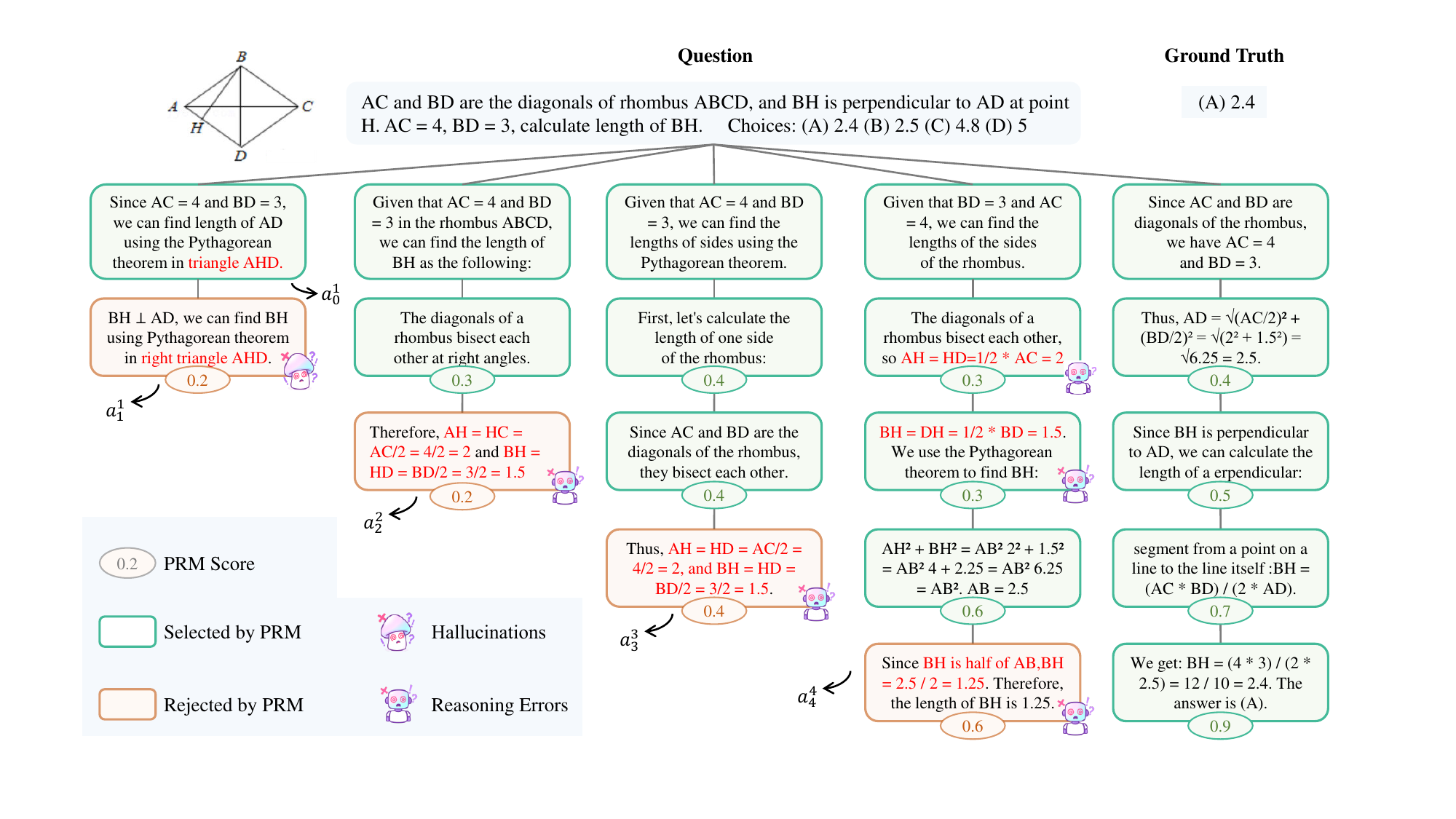}
	\vspace{-3mm}
	\caption{Qualitative case study on a geometry problem. PRM-BAS identifies and rejects steps with hallucinations and reasoning errors, guiding the policy model toward the correct conclusion.}
	\vspace{-3mm}
	\label{fig:case_study}
\end{figure*}

\subsection{Generalization of PRM-BAS \label{sec:general}}

\begin{table}[htb]
	\caption{Generalization of PRM-BAS to unseen policy models. \textcolor{gray}{Gray} values are reproduced results, which may differ from the original due to prompt or implementation differences.}
	\label{tab:general}
	\resizebox{1.0\linewidth}{!}{%
		
		\begin{tabular}{lccccc}
	\toprule
	\textbf{Method} & \textbf{\makecell{Math\\Vista}} & \textbf{\makecell{Math\\Vision}} & \textbf{\makecell{Chart\\QA}} & \textbf{M3CoT} & \textbf{AVG.} \\
	\midrule
	\multicolumn{6}{c}{\textit{Different Sizes}} \\
	Qwen2VL-2B & 43.0 & 12.4 & 73.5 & 45.0 & 43.5 \\
	\rowcolor{green!7}
	+ PRM-BAS & 59.7 & 18.3 & 79.4 & 61.7 & 54.8\raisebox{0.8ex}{\scriptsize\textcolor{green!60!black}{$\uparrow$ 11.3}} \\
	Qwen2.5VL-3B & 62.3 & 21.2 & 81.8 & 51.0 & 54.1 \\
	\rowcolor{green!7}
	+ PRM-BAS & 67.1 & 24.1 & 86.6 & 64.8 & 60.7\raisebox{0.8ex}{\scriptsize\textcolor{green!60!black}{$\uparrow$ 6.6}} \\
	\midrule
	\multicolumn{6}{c}{\textit{Different Series}} \\
	InternVL2-2B & 46.3 & 12.1 & \textcolor{gray}{67.6} & 47.7 & 43.4 \\
	\rowcolor{green!7}
	+ PRM-BAS & 53.6 & 17.4 & 73.8 & 58.7 & 50.9\raisebox{0.8ex}{\scriptsize\textcolor{green!60!black}{$\uparrow$ 7.5}} \\
	InternVL2-8B & 58.3 & 18.3 & \textcolor{gray}{79.6} & 59.3 & 53.9 \\	
	\rowcolor{green!7}
	+ PRM-BAS & 63.6 & 19.9 & 81.7 & 62.5 & 56.9\raisebox{0.8ex}{\scriptsize\textcolor{green!60!black}{$\uparrow$ 3.0}} \\
	\bottomrule
\end{tabular}
		
	}
\end{table}

\begin{table}[htb]
	
	\caption{Impact of training data source on PRM effectiveness with Qwen2VL-7B as the policy model, highlighting the policy-dependence issue.}
	\label{tab:dataset}
	\resizebox{0.99\linewidth}{!}{%

		\begin{tabular}{cc|ccccc}
	\toprule
	\multicolumn{2}{c|}{\textbf{Training Data}} 
	& \multirow{2}{*}{\centering \textbf{\makecell{Math\\Vista}}} 
	& \multirow{2}{*}{\centering \textbf{\makecell{Math\\Vision}}} 
	& \multirow{2}{*}{\centering \textbf{\makecell{Chart\\QA}}} 
	& \multirow{2}{*}{\centering \textbf{M3CoT}} 
	& \multirow{2}{*}{\centering \textbf{Avg.}} \\
	
	\cmidrule(lr){1-2}
	$\mathcal{D}_{\text{Qwen2}}$ & $\mathcal{D}_{\text{Qwen2.5}}$ 
	& & & & & \\
	
	\midrule
	\checkmark & & 67.2 & 22.8 & 86.0 & 73.1 & 62.3 \\
	& \checkmark & 65.2 & 23.4 & 85.4 & 70.4 & 61.1 \\
	\checkmark & \checkmark & 67.2 & 23.4 & 86.7 & 72.3 & 62.4 \\
	\bottomrule
\end{tabular}

	}
\end{table}

In constructing the PRM-BAS-300k training set, we use Qwen2-VL-7B and Qwen2.5-VL-7B as policy models. The resulting PRMs significantly improve the performance of these two models, as demonstrated in Section~\ref{sec:sota}. To evaluate generalization to unseen policy models, we further test the PRM on models of different sizes and architectures, including Qwen2VL-2B, Qwen2.5VL-3B, InternVL2-2B, and InternVL2-8B, none of which are involved in the training data construction. As shown in Table~\ref{tab:general}, PRM-BAS still yields performance gains, demonstrating a certain level of generalization. However, the improvements on the InternVL series are noticeably smaller compared to those on the Qwen series. This is because InternVL’s output style differs from Qwen's and is never seen during training. We refer to this as the \textit{Policy-Dependent Issue}, which has also been reported in prior work~\cite{liu2025can}.

To gain deeper insight into this issue, we conduct additional experiments, presented in Table \ref{tab:dataset}. Based on the policy model used during data construction, we split the full training set $\mathcal{D}$ into two subsets: $\mathcal{D}_{\text{Qwen2}}$ and $\mathcal{D}_{\text{Qwen2.5}}$. We then train separate PRMs on each subset and evaluate them using the same policy model, Qwen2-VL-7B. The results show that the PRM trained on $\mathcal{D}_{\text{Qwen2}}$ significantly outperforms the one trained on $\mathcal{D}_{\text{Qwen2.5}}$, suggesting that PRM performance is closely tied to the policy used during data construction. Furthermore, using only $\mathcal{D}_{\text{Qwen2}}$ achieves performance comparable to training on the full dataset $\mathcal{D}$, indicating that policy-aligned data plays a central role in effective PRM training.

\subsection{Case Study \label{sec:case_study}}

Figure~\ref{fig:case_study} presents a qualitative case study demonstrating how PRM-BAS operates on a geometry problem, which first requires interpreting the geometric elements in the image, and then applying relevant mathematical principles such as the properties of rhombuses, the triangle area formula, and the Pythagorean theorem. For clarity, we simplify the model's outputs without altering their original intent. In the first column, PRM correctly penalizes both $\boldsymbol{a}_0^1$ and $\boldsymbol{a}_1^1$. The policy model hallucinates that point AHD forms a triangle, despite the fact that all three points lie in a straight line. In the second column, the policy model incorrectly assumes that point H lies on the perpendicular bisector of segment AC, resulting in a low PRM score for $\boldsymbol{a}_2^2$. In the third column, the model prematurely draws a conclusion in $\boldsymbol{a}_3^3$ without sufficient supporting conditions, and PRM correctly rejects this action. A similar reasoning error occurs in the fourth column. These examples suggest that our PRM is capable of identifying both perception and reasoning errors, thereby guiding the policy model toward more accurate final answers.

\section{Limitation}
As discussed in Section~\ref{sec:general}, PRM-BAS exhibits a \textit{Policy-Dependent Issue}, where the effectiveness of the PRM is highly influenced by the consistency between the policy model used during training and the one used at inference. This issue can be mitigated by increasing the diversity of policy models during data construction, as prior work does \cite{visualprm,mulberry}. However, in real-world applications, this limitation is typically not a major concern, since the objective is to enhance the performance of a specific target policy rather than to achieve generalization across multiple policy models. For this reason, we do not increase the diversity of policy models during data construction.

Another limitation of this work is that, due to computational constraints, we do not experiment with larger models such as 32B or 72B. Nonetheless, we have evaluated PRM-BAS on multiple models ranging from 2B to 8B in size, showing encouraging generalization across different model sizes.

\vspace{-3mm}
\section{Conclusion}
In this paper, we present two key insights based on empirical analysis. (1) In the early stages of reasoning, the ground-truth reward variance is relatively high, indicating a large exploration space for MLLMs. (2) The training loss of the PRM is also higher in earlier steps, suggesting that the PRM struggles to accurately assess the quality of candidate actions when contextual information is limited. These observations motivate the design of a novel PRM-guided search strategy, PRM-BAS, which adopts a large beam size in early steps to improve tolerance to PRM prediction errors, and gradually reduces the beam size in later steps to enhance efficiency. In addition, we introduce a unified framework for training data construction and PRM learning.
Comprehensive evaluations on eight benchmarks confirm that PRM-BAS significantly enhances the reasoning performance of base policy models. Furthermore, we show that PRM-BAS generalizes well across models of varying sizes and architectures.
Our experiments demonstrate the effectiveness of using rank loss to capture relative action quality, and show that soft labels outperform hard labels, particularly that modern MLLMs exhibit a growing trend toward long-chain reasoning, where correct final answers may arise from incorrect intermediate steps via reflection and verification. 
Finally, we identify and analyze the \textit{Policy-Dependent Issue}, highlighting an important practical insight: the policy used during data collection should be aligned with the policy model used at inference time to maximize PRM effectiveness.

\FloatBarrier
\bibliographystyle{ACM-Reference-Format}
\bibliography{sample-base}


\begin{thebibliography}{64}


\ifx \showCODEN    \undefined \def \showCODEN     #1{\unskip}     \fi
\ifx \showDOI      \undefined \def \showDOI       #1{#1}\fi
\ifx \showISBNx    \undefined \def \showISBNx     #1{\unskip}     \fi
\ifx \showISBNxiii \undefined \def \showISBNxiii  #1{\unskip}     \fi
\ifx \showISSN     \undefined \def \showISSN      #1{\unskip}     \fi
\ifx \showLCCN     \undefined \def \showLCCN      #1{\unskip}     \fi
\ifx \shownote     \undefined \def \shownote      #1{#1}          \fi
\ifx \showarticletitle \undefined \def \showarticletitle #1{#1}   \fi
\ifx \showURL      \undefined \def \showURL       {\relax}        \fi
\providecommand\bibfield[2]{#2}
\providecommand\bibinfo[2]{#2}
\providecommand\natexlab[1]{#1}
\providecommand\showeprint[2][]{arXiv:#2}

\bibitem[Anthropic(2024)]%
        {anthropic2024claude}
\bibfield{author}{\bibinfo{person}{Anthropic}.}
  \bibinfo{year}{2024}\natexlab{}.
\newblock \bibinfo{title}{Claude 3.5 Sonnet}.
\newblock
\newblock
\urldef\tempurl%
\url{https://www.anthropic.com/news/claude-3-5-sonnet}
\showURL{%
\tempurl}
\newblock
\shownote{Accessed: 2025-04-10}.


\bibitem[Bai et~al\mbox{.}(2025)]%
        {qwen2.5vl}
\bibfield{author}{\bibinfo{person}{Shuai Bai}, \bibinfo{person}{Keqin Chen},
  \bibinfo{person}{Xuejing Liu}, \bibinfo{person}{Jialin Wang},
  \bibinfo{person}{Wenbin Ge}, \bibinfo{person}{Sibo Song},
  \bibinfo{person}{Kai Dang}, \bibinfo{person}{Peng Wang},
  \bibinfo{person}{Shijie Wang}, \bibinfo{person}{Jun Tang},
  \bibinfo{person}{Humen Zhong}, \bibinfo{person}{Yuanzhi Zhu},
  \bibinfo{person}{Mingkun Yang}, \bibinfo{person}{Zhaohai Li},
  \bibinfo{person}{Jianqiang Wan}, \bibinfo{person}{Pengfei Wang},
  \bibinfo{person}{Wei Ding}, \bibinfo{person}{Zheren Fu},
  \bibinfo{person}{Yiheng Xu}, \bibinfo{person}{Jiabo Ye}, \bibinfo{person}{Xi
  Zhang}, \bibinfo{person}{Tianbao Xie}, \bibinfo{person}{Zesen Cheng},
  \bibinfo{person}{Hang Zhang}, \bibinfo{person}{Zhibo Yang},
  \bibinfo{person}{Haiyang Xu}, {and} \bibinfo{person}{Junyang Lin}.}
  \bibinfo{year}{2025}\natexlab{}.
\newblock \showarticletitle{Qwen2.5-VL Technical Report}.
\newblock \bibinfo{journal}{\emph{arXiv preprint arXiv:2502.13923}}
  (\bibinfo{year}{2025}).
\newblock


\bibitem[Chen et~al\mbox{.}(2024a)]%
        {chen2024alphamath}
\bibfield{author}{\bibinfo{person}{Guoxin Chen}, \bibinfo{person}{Minpeng
  Liao}, \bibinfo{person}{Chengxi Li}, {and} \bibinfo{person}{Kai Fan}.}
  \bibinfo{year}{2024}\natexlab{a}.
\newblock \showarticletitle{Alphamath almost zero: process supervision without
  process}.
\newblock \bibinfo{journal}{\emph{arXiv preprint arXiv:2405.03553}}
  (\bibinfo{year}{2024}).
\newblock


\bibitem[Chen et~al\mbox{.}(2025)]%
        {chen2025r1v}
\bibfield{author}{\bibinfo{person}{Liang Chen}, \bibinfo{person}{Lei Li},
  \bibinfo{person}{Haozhe Zhao}, \bibinfo{person}{Yifan Song}, {and}
  \bibinfo{person}{Vinci}.} \bibinfo{year}{2025}\natexlab{}.
\newblock \bibinfo{title}{R1-V: Reinforcing Super Generalization Ability in
  Vision-Language Models with Less Than \$3}.
\newblock \bibinfo{howpublished}{\url{https://github.com/Deep-Agent/R1-V}}.
\newblock
\newblock
\shownote{Accessed: 2025-02-02}.


\bibitem[Chen et~al\mbox{.}(2024b)]%
        {m3cot}
\bibfield{author}{\bibinfo{person}{Qiguang Chen}, \bibinfo{person}{Libo Qin},
  \bibinfo{person}{Jin Zhang}, \bibinfo{person}{Zhi Chen},
  \bibinfo{person}{Xiao Xu}, {and} \bibinfo{person}{Wanxiang Che}.}
  \bibinfo{year}{2024}\natexlab{b}.
\newblock \showarticletitle{M3CoT: A Novel Benchmark for Multi-Domain
  Multi-step Multi-modal Chain-of-Thought}. In
  \bibinfo{booktitle}{\emph{Proceedings of the 62nd Annual Meeting of the
  Association for Computational Linguistics}}. \bibinfo{pages}{8199--8221}.
\newblock


\bibitem[Chen et~al\mbox{.}(2024c)]%
        {chen2024expanding}
\bibfield{author}{\bibinfo{person}{Zhe Chen}, \bibinfo{person}{Weiyun Wang},
  \bibinfo{person}{Yue Cao}, \bibinfo{person}{Yangzhou Liu},
  \bibinfo{person}{Zhangwei Gao}, \bibinfo{person}{Erfei Cui},
  \bibinfo{person}{Jinguo Zhu}, \bibinfo{person}{Shenglong Ye},
  \bibinfo{person}{Hao Tian}, \bibinfo{person}{Zhaoyang Liu}, {et~al\mbox{.}}}
  \bibinfo{year}{2024}\natexlab{c}.
\newblock \showarticletitle{Expanding performance boundaries of open-source
  multimodal models with model, data, and test-time scaling}.
\newblock \bibinfo{journal}{\emph{arXiv preprint arXiv:2412.05271}}
  (\bibinfo{year}{2024}).
\newblock


\bibitem[Chen et~al\mbox{.}(2024d)]%
        {chen2024far}
\bibfield{author}{\bibinfo{person}{Zhe Chen}, \bibinfo{person}{Weiyun Wang},
  \bibinfo{person}{Hao Tian}, \bibinfo{person}{Shenglong Ye},
  \bibinfo{person}{Zhangwei Gao}, \bibinfo{person}{Erfei Cui},
  \bibinfo{person}{Wenwen Tong}, \bibinfo{person}{Kongzhi Hu},
  \bibinfo{person}{Jiapeng Luo}, \bibinfo{person}{Zheng Ma}, {et~al\mbox{.}}}
  \bibinfo{year}{2024}\natexlab{d}.
\newblock \showarticletitle{How far are we to gpt-4v? closing the gap to
  commercial multimodal models with open-source suites}.
\newblock \bibinfo{journal}{\emph{Science China Information Sciences}}
  \bibinfo{volume}{67}, \bibinfo{number}{12} (\bibinfo{year}{2024}),
  \bibinfo{pages}{220101}.
\newblock


\bibitem[Cheng et~al\mbox{.}(2024)]%
        {cheng2024vision}
\bibfield{author}{\bibinfo{person}{Kanzhi Cheng}, \bibinfo{person}{Yantao Li},
  \bibinfo{person}{Fangzhi Xu}, \bibinfo{person}{Jianbing Zhang},
  \bibinfo{person}{Hao Zhou}, {and} \bibinfo{person}{Yang Liu}.}
  \bibinfo{year}{2024}\natexlab{}.
\newblock \showarticletitle{Vision-language models can self-improve reasoning
  via reflection}.
\newblock \bibinfo{journal}{\emph{arXiv preprint arXiv:2411.00855}}
  (\bibinfo{year}{2024}).
\newblock


\bibitem[Dong et~al\mbox{.}(2024c)]%
        {dong2024rlhf}
\bibfield{author}{\bibinfo{person}{Hanze Dong}, \bibinfo{person}{Wei Xiong},
  \bibinfo{person}{Bo Pang}, \bibinfo{person}{Haoxiang Wang},
  \bibinfo{person}{Han Zhao}, \bibinfo{person}{Yingbo Zhou},
  \bibinfo{person}{Nan Jiang}, \bibinfo{person}{Doyen Sahoo},
  \bibinfo{person}{Caiming Xiong}, {and} \bibinfo{person}{Tong Zhang}.}
  \bibinfo{year}{2024}\natexlab{c}.
\newblock \showarticletitle{Rlhf workflow: From reward modeling to online
  rlhf}.
\newblock \bibinfo{journal}{\emph{arXiv preprint arXiv:2405.07863}}
  (\bibinfo{year}{2024}).
\newblock


\bibitem[Dong et~al\mbox{.}(2024a)]%
        {insight-v}
\bibfield{author}{\bibinfo{person}{Yuhao Dong}, \bibinfo{person}{Zuyan Liu},
  \bibinfo{person}{Hai-Long Sun}, \bibinfo{person}{Jingkang Yang},
  \bibinfo{person}{Winston Hu}, \bibinfo{person}{Yongming Rao}, {and}
  \bibinfo{person}{Ziwei Liu}.} \bibinfo{year}{2024}\natexlab{a}.
\newblock \showarticletitle{Insight-v: Exploring long-chain visual reasoning
  with multimodal large language models}.
\newblock \bibinfo{journal}{\emph{arXiv preprint arXiv:2411.14432}}
  (\bibinfo{year}{2024}).
\newblock


\bibitem[Dong et~al\mbox{.}(2024b)]%
        {dong2024insight}
\bibfield{author}{\bibinfo{person}{Yuhao Dong}, \bibinfo{person}{Zuyan Liu},
  \bibinfo{person}{Hai-Long Sun}, \bibinfo{person}{Jingkang Yang},
  \bibinfo{person}{Winston Hu}, \bibinfo{person}{Yongming Rao}, {and}
  \bibinfo{person}{Ziwei Liu}.} \bibinfo{year}{2024}\natexlab{b}.
\newblock \showarticletitle{Insight-v: Exploring long-chain visual reasoning
  with multimodal large language models}.
\newblock \bibinfo{journal}{\emph{arXiv preprint arXiv:2411.14432}}
  (\bibinfo{year}{2024}).
\newblock


\bibitem[Du et~al\mbox{.}(2025)]%
        {virgo}
\bibfield{author}{\bibinfo{person}{Yifan Du}, \bibinfo{person}{Zikang Liu},
  \bibinfo{person}{Yifan Li}, \bibinfo{person}{Wayne~Xin Zhao},
  \bibinfo{person}{Yuqi Huo}, \bibinfo{person}{Bingning Wang},
  \bibinfo{person}{Weipeng Chen}, \bibinfo{person}{Zheng Liu},
  \bibinfo{person}{Zhongyuan Wang}, {and} \bibinfo{person}{Ji-Rong Wen}.}
  \bibinfo{year}{2025}\natexlab{}.
\newblock \showarticletitle{Virgo: A Preliminary Exploration on Reproducing
  o1-like MLLM}.
\newblock \bibinfo{journal}{\emph{arXiv preprint arXiv:2501.01904}}
  (\bibinfo{year}{2025}).
\newblock


\bibitem[Gao et~al\mbox{.}(2024)]%
        {cantor}
\bibfield{author}{\bibinfo{person}{Timin Gao}, \bibinfo{person}{Peixian Chen},
  \bibinfo{person}{Mengdan Zhang}, \bibinfo{person}{Chaoyou Fu},
  \bibinfo{person}{Yunhang Shen}, \bibinfo{person}{Yan Zhang},
  \bibinfo{person}{Shengchuan Zhang}, \bibinfo{person}{Xiawu Zheng},
  \bibinfo{person}{Xing Sun}, \bibinfo{person}{Liujuan Cao}, {et~al\mbox{.}}}
  \bibinfo{year}{2024}\natexlab{}.
\newblock \showarticletitle{Cantor: Inspiring multimodal chain-of-thought of
  mllm}. In \bibinfo{booktitle}{\emph{Proceedings of the 32nd ACM International
  Conference on Multimedia}}. \bibinfo{pages}{9096--9105}.
\newblock


\bibitem[Guo et~al\mbox{.}(2025)]%
        {ds-r1}
\bibfield{author}{\bibinfo{person}{Daya Guo}, \bibinfo{person}{Dejian Yang},
  \bibinfo{person}{Haowei Zhang}, \bibinfo{person}{Junxiao Song},
  \bibinfo{person}{Ruoyu Zhang}, \bibinfo{person}{Runxin Xu},
  \bibinfo{person}{Qihao Zhu}, \bibinfo{person}{Shirong Ma},
  \bibinfo{person}{Peiyi Wang}, \bibinfo{person}{Xiao Bi}, {et~al\mbox{.}}}
  \bibinfo{year}{2025}\natexlab{}.
\newblock \showarticletitle{Deepseek-r1: Incentivizing reasoning capability in
  llms via reinforcement learning}.
\newblock \bibinfo{journal}{\emph{arXiv preprint arXiv:2501.12948}}
  (\bibinfo{year}{2025}).
\newblock


\bibitem[Huang et~al\mbox{.}(2025)]%
        {vision-r1}
\bibfield{author}{\bibinfo{person}{Wenxuan Huang}, \bibinfo{person}{Bohan Jia},
  \bibinfo{person}{Zijie Zhai}, \bibinfo{person}{Shaosheng Cao},
  \bibinfo{person}{Zheyu Ye}, \bibinfo{person}{Fei Zhao}, \bibinfo{person}{Yao
  Hu}, {and} \bibinfo{person}{Shaohui Lin}.} \bibinfo{year}{2025}\natexlab{}.
\newblock \showarticletitle{Vision-r1: Incentivizing reasoning capability in
  multimodal large language models}.
\newblock \bibinfo{journal}{\emph{arXiv preprint arXiv:2503.06749}}
  (\bibinfo{year}{2025}).
\newblock


\bibitem[Hurst et~al\mbox{.}(2024)]%
        {gpt4o}
\bibfield{author}{\bibinfo{person}{Aaron Hurst}, \bibinfo{person}{Adam Lerer},
  \bibinfo{person}{Adam~P Goucher}, \bibinfo{person}{Adam Perelman},
  \bibinfo{person}{Aditya Ramesh}, \bibinfo{person}{Aidan Clark},
  \bibinfo{person}{AJ Ostrow}, \bibinfo{person}{Akila Welihinda},
  \bibinfo{person}{Alan Hayes}, \bibinfo{person}{Alec Radford},
  {et~al\mbox{.}}} \bibinfo{year}{2024}\natexlab{}.
\newblock \showarticletitle{Gpt-4o system card}.
\newblock \bibinfo{journal}{\emph{arXiv preprint arXiv:2410.21276}}
  (\bibinfo{year}{2024}).
\newblock


\bibitem[Jaech et~al\mbox{.}(2024)]%
        {o1}
\bibfield{author}{\bibinfo{person}{Aaron Jaech}, \bibinfo{person}{Adam Kalai},
  \bibinfo{person}{Adam Lerer}, \bibinfo{person}{Adam Richardson},
  \bibinfo{person}{Ahmed El-Kishky}, \bibinfo{person}{Aiden Low},
  \bibinfo{person}{Alec Helyar}, \bibinfo{person}{Aleksander Madry},
  \bibinfo{person}{Alex Beutel}, \bibinfo{person}{Alex Carney},
  {et~al\mbox{.}}} \bibinfo{year}{2024}\natexlab{}.
\newblock \showarticletitle{Openai o1 system card}.
\newblock \bibinfo{journal}{\emph{arXiv preprint arXiv:2412.16720}}
  (\bibinfo{year}{2024}).
\newblock


\bibitem[Lightman et~al\mbox{.}(2023)]%
        {lightman2023let}
\bibfield{author}{\bibinfo{person}{Hunter Lightman}, \bibinfo{person}{Vineet
  Kosaraju}, \bibinfo{person}{Yuri Burda}, \bibinfo{person}{Harrison Edwards},
  \bibinfo{person}{Bowen Baker}, \bibinfo{person}{Teddy Lee},
  \bibinfo{person}{Jan Leike}, \bibinfo{person}{John Schulman},
  \bibinfo{person}{Ilya Sutskever}, {and} \bibinfo{person}{Karl Cobbe}.}
  \bibinfo{year}{2023}\natexlab{}.
\newblock \showarticletitle{Let's verify step by step}. In
  \bibinfo{booktitle}{\emph{The Twelfth International Conference on Learning
  Representations}}.
\newblock


\bibitem[Lin et~al\mbox{.}(2025)]%
        {lin2025mind}
\bibfield{author}{\bibinfo{person}{Zhiyu Lin}, \bibinfo{person}{Yifei Gao},
  \bibinfo{person}{Xian Zhao}, \bibinfo{person}{Yunfan Yang}, {and}
  \bibinfo{person}{Jitao Sang}.} \bibinfo{year}{2025}\natexlab{}.
\newblock \showarticletitle{Mind with Eyes: from Language Reasoning to
  Multimodal Reasoning}.
\newblock \bibinfo{journal}{\emph{arXiv preprint arXiv:2503.18071}}
  (\bibinfo{year}{2025}).
\newblock


\bibitem[Lin et~al\mbox{.}(2024)]%
        {lin2024understanding}
\bibfield{author}{\bibinfo{person}{Zhutian Lin}, \bibinfo{person}{Junwei Pan},
  \bibinfo{person}{Shangyu Zhang}, \bibinfo{person}{Ximei Wang},
  \bibinfo{person}{Xi Xiao}, \bibinfo{person}{Shudong Huang},
  \bibinfo{person}{Lei Xiao}, {and} \bibinfo{person}{Jie Jiang}.}
  \bibinfo{year}{2024}\natexlab{}.
\newblock \showarticletitle{Understanding the Ranking Loss for Recommendation
  with Sparse User Feedback}. In \bibinfo{booktitle}{\emph{Proceedings of the
  30th ACM SIGKDD Conference on Knowledge Discovery and Data Mining}}.
  \bibinfo{pages}{5409--5418}.
\newblock


\bibitem[Liu et~al\mbox{.}(2024)]%
        {liu2024llavanext}
\bibfield{author}{\bibinfo{person}{Haotian Liu}, \bibinfo{person}{Chunyuan Li},
  \bibinfo{person}{Yuheng Li}, \bibinfo{person}{Bo Li},
  \bibinfo{person}{Yuanhan Zhang}, \bibinfo{person}{Sheng Shen}, {and}
  \bibinfo{person}{Yong~Jae Lee}.} \bibinfo{year}{2024}\natexlab{}.
\newblock \bibinfo{title}{LLaVA-NeXT: Improved reasoning, OCR, and world
  knowledge}.
\newblock
\newblock
\urldef\tempurl%
\url{https://llava-vl.github.io/blog/2024-01-30-llava-next/}
\showURL{%
\tempurl}


\bibitem[Liu et~al\mbox{.}(2025a)]%
        {liu2025can}
\bibfield{author}{\bibinfo{person}{Runze Liu}, \bibinfo{person}{Junqi Gao},
  \bibinfo{person}{Jian Zhao}, \bibinfo{person}{Kaiyan Zhang},
  \bibinfo{person}{Xiu Li}, \bibinfo{person}{Biqing Qi}, \bibinfo{person}{Wanli
  Ouyang}, {and} \bibinfo{person}{Bowen Zhou}.}
  \bibinfo{year}{2025}\natexlab{a}.
\newblock \showarticletitle{Can 1B LLM Surpass 405B LLM? Rethinking
  Compute-Optimal Test-Time Scaling}.
\newblock \bibinfo{journal}{\emph{arXiv preprint arXiv:2502.06703}}
  (\bibinfo{year}{2025}).
\newblock


\bibitem[Liu et~al\mbox{.}(2025b)]%
        {visual-rft}
\bibfield{author}{\bibinfo{person}{Ziyu Liu}, \bibinfo{person}{Zeyi Sun},
  \bibinfo{person}{Yuhang Zang}, \bibinfo{person}{Xiaoyi Dong},
  \bibinfo{person}{Yuhang Cao}, \bibinfo{person}{Haodong Duan},
  \bibinfo{person}{Dahua Lin}, {and} \bibinfo{person}{Jiaqi Wang}.}
  \bibinfo{year}{2025}\natexlab{b}.
\newblock \showarticletitle{Visual-rft: Visual reinforcement fine-tuning}.
\newblock \bibinfo{journal}{\emph{arXiv preprint arXiv:2503.01785}}
  (\bibinfo{year}{2025}).
\newblock


\bibitem[Liu et~al\mbox{.}(2025c)]%
        {liu2025inference}
\bibfield{author}{\bibinfo{person}{Zijun Liu}, \bibinfo{person}{Peiyi Wang},
  \bibinfo{person}{Runxin Xu}, \bibinfo{person}{Shirong Ma},
  \bibinfo{person}{Chong Ruan}, \bibinfo{person}{Peng Li},
  \bibinfo{person}{Yang Liu}, {and} \bibinfo{person}{Yu Wu}.}
  \bibinfo{year}{2025}\natexlab{c}.
\newblock \showarticletitle{Inference-Time Scaling for Generalist Reward
  Modeling}.
\newblock \bibinfo{journal}{\emph{arXiv preprint arXiv:2504.02495}}
  (\bibinfo{year}{2025}).
\newblock


\bibitem[Loshchilov and Hutter(2019)]%
        {adamw}
\bibfield{author}{\bibinfo{person}{Ilya Loshchilov} {and}
  \bibinfo{person}{Frank Hutter}.} \bibinfo{year}{2019}\natexlab{}.
\newblock \showarticletitle{Decoupled Weight Decay Regularization}. In
  \bibinfo{booktitle}{\emph{7th International Conference on Learning
  Representations}}.
\newblock


\bibitem[Lu et~al\mbox{.}(2024b)]%
        {lu2024deepseek}
\bibfield{author}{\bibinfo{person}{Haoyu Lu}, \bibinfo{person}{Wen Liu},
  \bibinfo{person}{Bo Zhang}, \bibinfo{person}{Bingxuan Wang},
  \bibinfo{person}{Kai Dong}, \bibinfo{person}{Bo Liu},
  \bibinfo{person}{Jingxiang Sun}, \bibinfo{person}{Tongzheng Ren},
  \bibinfo{person}{Zhuoshu Li}, \bibinfo{person}{Hao Yang}, {et~al\mbox{.}}}
  \bibinfo{year}{2024}\natexlab{b}.
\newblock \showarticletitle{Deepseek-vl: towards real-world vision-language
  understanding}.
\newblock \bibinfo{journal}{\emph{arXiv preprint arXiv:2403.05525}}
  (\bibinfo{year}{2024}).
\newblock


\bibitem[Lu et~al\mbox{.}(2024a)]%
        {mathvista}
\bibfield{author}{\bibinfo{person}{Pan Lu}, \bibinfo{person}{Hritik Bansal},
  \bibinfo{person}{Tony Xia}, \bibinfo{person}{Jiacheng Liu},
  \bibinfo{person}{Chunyuan Li}, \bibinfo{person}{Hannaneh Hajishirzi},
  \bibinfo{person}{Hao Cheng}, \bibinfo{person}{Kai-Wei Chang},
  \bibinfo{person}{Michel Galley}, {and} \bibinfo{person}{Jianfeng Gao}.}
  \bibinfo{year}{2024}\natexlab{a}.
\newblock \showarticletitle{MathVista: Evaluating Mathematical Reasoning of
  Foundation Models in Visual Contexts}. In
  \bibinfo{booktitle}{\emph{International Conference on Learning
  Representations (ICLR)}}.
\newblock


\bibitem[Luo et~al\mbox{.}(2024)]%
        {luo2024improve}
\bibfield{author}{\bibinfo{person}{Liangchen Luo}, \bibinfo{person}{Yinxiao
  Liu}, \bibinfo{person}{Rosanne Liu}, \bibinfo{person}{Samrat Phatale},
  \bibinfo{person}{Harsh Lara}, \bibinfo{person}{Yunxuan Li},
  \bibinfo{person}{Lei Shu}, \bibinfo{person}{Yun Zhu}, \bibinfo{person}{Lei
  Meng}, \bibinfo{person}{Jiao Sun}, {et~al\mbox{.}}}
  \bibinfo{year}{2024}\natexlab{}.
\newblock \showarticletitle{Improve mathematical reasoning in language models
  by automated process supervision}.
\newblock \bibinfo{journal}{\emph{arXiv preprint arXiv:2406.06592}}
  \bibinfo{volume}{2} (\bibinfo{year}{2024}).
\newblock


\bibitem[Masry et~al\mbox{.}(2022)]%
        {chartqa}
\bibfield{author}{\bibinfo{person}{Ahmed Masry}, \bibinfo{person}{Xuan~Long
  Do}, \bibinfo{person}{Jia~Qing Tan}, \bibinfo{person}{Shafiq Joty}, {and}
  \bibinfo{person}{Enamul Hoque}.} \bibinfo{year}{2022}\natexlab{}.
\newblock \showarticletitle{ChartQA: A Benchmark for Question Answering about
  Charts with Visual and Logical Reasoning}. In
  \bibinfo{booktitle}{\emph{Findings of the Association for Computational
  Linguistics: ACL 2022}}. \bibinfo{pages}{2263--2279}.
\newblock


\bibitem[Mitra et~al\mbox{.}(2024)]%
        {ccot}
\bibfield{author}{\bibinfo{person}{Chancharik Mitra}, \bibinfo{person}{Brandon
  Huang}, \bibinfo{person}{Trevor Darrell}, {and} \bibinfo{person}{Roei
  Herzig}.} \bibinfo{year}{2024}\natexlab{}.
\newblock \showarticletitle{Compositional chain-of-thought prompting for large
  multimodal models}. In \bibinfo{booktitle}{\emph{Proceedings of the IEEE/CVF
  Conference on Computer Vision and Pattern Recognition}}.
  \bibinfo{pages}{14420--14431}.
\newblock


\bibitem[Rajbhandari et~al\mbox{.}(2020)]%
        {zeromemory}
\bibfield{author}{\bibinfo{person}{Samyam Rajbhandari}, \bibinfo{person}{Jeff
  Rasley}, \bibinfo{person}{Olatunji Ruwase}, {and} \bibinfo{person}{Yuxiong
  He}.} \bibinfo{year}{2020}\natexlab{}.
\newblock \showarticletitle{Zero: Memory optimizations toward training trillion
  parameter models}. In \bibinfo{booktitle}{\emph{SC20: International
  Conference for High Performance Computing, Networking, Storage and
  Analysis}}. IEEE, \bibinfo{pages}{1--16}.
\newblock


\bibitem[Saikh et~al\mbox{.}(2022)]%
        {saikh2022scienceqa}
\bibfield{author}{\bibinfo{person}{Tanik Saikh}, \bibinfo{person}{Tirthankar
  Ghosal}, \bibinfo{person}{Amish Mittal}, \bibinfo{person}{Asif Ekbal}, {and}
  \bibinfo{person}{Pushpak Bhattacharyya}.} \bibinfo{year}{2022}\natexlab{}.
\newblock \showarticletitle{Scienceqa: A novel resource for question answering
  on scholarly articles}.
\newblock \bibinfo{journal}{\emph{International Journal on Digital Libraries}}
  \bibinfo{volume}{23}, \bibinfo{number}{3} (\bibinfo{year}{2022}),
  \bibinfo{pages}{289--301}.
\newblock


\bibitem[Shao et~al\mbox{.}(2024)]%
        {shao2024visual}
\bibfield{author}{\bibinfo{person}{Hao Shao}, \bibinfo{person}{Shengju Qian},
  \bibinfo{person}{Han Xiao}, \bibinfo{person}{Guanglu Song},
  \bibinfo{person}{Zhuofan Zong}, \bibinfo{person}{Letian Wang},
  \bibinfo{person}{Yu Liu}, {and} \bibinfo{person}{Hongsheng Li}.}
  \bibinfo{year}{2024}\natexlab{}.
\newblock \showarticletitle{Visual cot: Advancing multi-modal language models
  with a comprehensive dataset and benchmark for chain-of-thought reasoning}.
\newblock \bibinfo{journal}{\emph{Advances in Neural Information Processing
  Systems}}  \bibinfo{volume}{37} (\bibinfo{year}{2024}),
  \bibinfo{pages}{8612--8642}.
\newblock


\bibitem[Sheng et~al\mbox{.}(2023)]%
        {sheng2023joint}
\bibfield{author}{\bibinfo{person}{Xiang-Rong Sheng}, \bibinfo{person}{Jingyue
  Gao}, \bibinfo{person}{Yueyao Cheng}, \bibinfo{person}{Siran Yang},
  \bibinfo{person}{Shuguang Han}, \bibinfo{person}{Hongbo Deng},
  \bibinfo{person}{Yuning Jiang}, \bibinfo{person}{Jian Xu}, {and}
  \bibinfo{person}{Bo Zheng}.} \bibinfo{year}{2023}\natexlab{}.
\newblock \showarticletitle{Joint optimization of ranking and calibration with
  contextualized hybrid model}. In \bibinfo{booktitle}{\emph{Proceedings of the
  29th ACM SIGKDD Conference on Knowledge Discovery and Data Mining}}.
  \bibinfo{pages}{4813--4822}.
\newblock


\bibitem[Shi et~al\mbox{.}(2024)]%
        {shi2024math}
\bibfield{author}{\bibinfo{person}{Wenhao Shi}, \bibinfo{person}{Zhiqiang Hu},
  \bibinfo{person}{Yi Bin}, \bibinfo{person}{Junhua Liu}, \bibinfo{person}{Yang
  Yang}, \bibinfo{person}{See-Kiong Ng}, \bibinfo{person}{Lidong Bing}, {and}
  \bibinfo{person}{Roy Ka-Wei Lee}.} \bibinfo{year}{2024}\natexlab{}.
\newblock \showarticletitle{Math-llava: Bootstrapping mathematical reasoning
  for multimodal large language models}.
\newblock \bibinfo{journal}{\emph{arXiv preprint arXiv:2406.17294}}
  (\bibinfo{year}{2024}).
\newblock


\bibitem[Snell et~al\mbox{.}(2024)]%
        {snell2024scaling}
\bibfield{author}{\bibinfo{person}{Charlie Snell}, \bibinfo{person}{Jaehoon
  Lee}, \bibinfo{person}{Kelvin Xu}, {and} \bibinfo{person}{Aviral Kumar}.}
  \bibinfo{year}{2024}\natexlab{}.
\newblock \showarticletitle{Scaling llm test-time compute optimally can be more
  effective than scaling model parameters}.
\newblock \bibinfo{journal}{\emph{arXiv preprint arXiv:2408.03314}}
  (\bibinfo{year}{2024}).
\newblock


\bibitem[Team et~al\mbox{.}(2024)]%
        {team2024gemini}
\bibfield{author}{\bibinfo{person}{Gemini Team}, \bibinfo{person}{Petko
  Georgiev}, \bibinfo{person}{Ving~Ian Lei}, \bibinfo{person}{Ryan Burnell},
  \bibinfo{person}{Libin Bai}, \bibinfo{person}{Anmol Gulati},
  \bibinfo{person}{Garrett Tanzer}, \bibinfo{person}{Damien Vincent},
  \bibinfo{person}{Zhufeng Pan}, \bibinfo{person}{Shibo Wang}, {et~al\mbox{.}}}
  \bibinfo{year}{2024}\natexlab{}.
\newblock \showarticletitle{Gemini 1.5: Unlocking multimodal understanding
  across millions of tokens of context}.
\newblock \bibinfo{journal}{\emph{arXiv preprint arXiv:2403.05530}}
  (\bibinfo{year}{2024}).
\newblock


\bibitem[Team(2024)]%
        {qvq}
\bibfield{author}{\bibinfo{person}{Qwen Team}.}
  \bibinfo{year}{2024}\natexlab{}.
\newblock \showarticletitle{Qvq: To see the world with wisdom}.
\newblock  (\bibinfo{year}{2024}).
\newblock


\bibitem[Thawakar et~al\mbox{.}(2025)]%
        {thawakar2025llamav}
\bibfield{author}{\bibinfo{person}{Omkar Thawakar}, \bibinfo{person}{Dinura
  Dissanayake}, \bibinfo{person}{Ketan More}, \bibinfo{person}{Ritesh Thawkar},
  \bibinfo{person}{Ahmed Heakl}, \bibinfo{person}{Noor Ahsan},
  \bibinfo{person}{Yuhao Li}, \bibinfo{person}{Mohammed Zumri},
  \bibinfo{person}{Jean Lahoud}, \bibinfo{person}{Rao~Muhammad Anwer},
  {et~al\mbox{.}}} \bibinfo{year}{2025}\natexlab{}.
\newblock \showarticletitle{Llamav-o1: Rethinking step-by-step visual reasoning
  in llms}.
\newblock \bibinfo{journal}{\emph{arXiv preprint arXiv:2501.06186}}
  (\bibinfo{year}{2025}).
\newblock


\bibitem[Uesato et~al\mbox{.}(2022)]%
        {uesato2022solving}
\bibfield{author}{\bibinfo{person}{Jonathan Uesato}, \bibinfo{person}{Nate
  Kushman}, \bibinfo{person}{Ramana Kumar}, \bibinfo{person}{Francis Song},
  \bibinfo{person}{Noah Siegel}, \bibinfo{person}{Lisa Wang},
  \bibinfo{person}{Antonia Creswell}, \bibinfo{person}{Geoffrey Irving}, {and}
  \bibinfo{person}{Irina Higgins}.} \bibinfo{year}{2022}\natexlab{}.
\newblock \showarticletitle{Solving math word problems with process-and
  outcome-based feedback}.
\newblock \bibinfo{journal}{\emph{arXiv preprint arXiv:2211.14275}}
  (\bibinfo{year}{2022}).
\newblock


\bibitem[Wan et~al\mbox{.}(2024)]%
        {wan2024alphazero}
\bibfield{author}{\bibinfo{person}{Ziyu Wan}, \bibinfo{person}{Xidong Feng},
  \bibinfo{person}{Muning Wen}, \bibinfo{person}{Stephen~Marcus McAleer},
  \bibinfo{person}{Ying Wen}, \bibinfo{person}{Weinan Zhang}, {and}
  \bibinfo{person}{Jun Wang}.} \bibinfo{year}{2024}\natexlab{}.
\newblock \showarticletitle{Alphazero-like tree-search can guide large language
  model decoding and training}. In \bibinfo{booktitle}{\emph{Forty-first
  International Conference on Machine Learning}}.
\newblock


\bibitem[Wang et~al\mbox{.}(2024e)]%
        {wang2024interpretable}
\bibfield{author}{\bibinfo{person}{Haoxiang Wang}, \bibinfo{person}{Wei Xiong},
  \bibinfo{person}{Tengyang Xie}, \bibinfo{person}{Han Zhao}, {and}
  \bibinfo{person}{Tong Zhang}.} \bibinfo{year}{2024}\natexlab{e}.
\newblock \showarticletitle{Interpretable preferences via multi-objective
  reward modeling and mixture-of-experts}.
\newblock \bibinfo{journal}{\emph{arXiv preprint arXiv:2406.12845}}
  (\bibinfo{year}{2024}).
\newblock


\bibitem[Wang et~al\mbox{.}(2024d)]%
        {mathvision}
\bibfield{author}{\bibinfo{person}{Ke Wang}, \bibinfo{person}{Junting Pan},
  \bibinfo{person}{Weikang Shi}, \bibinfo{person}{Zimu Lu},
  \bibinfo{person}{Houxing Ren}, \bibinfo{person}{Aojun Zhou},
  \bibinfo{person}{Mingjie Zhan}, {and} \bibinfo{person}{Hongsheng Li}.}
  \bibinfo{year}{2024}\natexlab{d}.
\newblock \showarticletitle{Measuring Multimodal Mathematical Reasoning with
  MATH-Vision Dataset}. In \bibinfo{booktitle}{\emph{The Thirty-eight
  Conference on Neural Information Processing Systems Datasets and Benchmarks
  Track}}.
\newblock


\bibitem[Wang et~al\mbox{.}(2024a)]%
        {qwen2vl}
\bibfield{author}{\bibinfo{person}{Peng Wang}, \bibinfo{person}{Shuai Bai},
  \bibinfo{person}{Sinan Tan}, \bibinfo{person}{Shijie Wang},
  \bibinfo{person}{Zhihao Fan}, \bibinfo{person}{Jinze Bai},
  \bibinfo{person}{Keqin Chen}, \bibinfo{person}{Xuejing Liu},
  \bibinfo{person}{Jialin Wang}, \bibinfo{person}{Wenbin Ge},
  \bibinfo{person}{Yang Fan}, \bibinfo{person}{Kai Dang},
  \bibinfo{person}{Mengfei Du}, \bibinfo{person}{Xuancheng Ren},
  \bibinfo{person}{Rui Men}, \bibinfo{person}{Dayiheng Liu},
  \bibinfo{person}{Chang Zhou}, \bibinfo{person}{Jingren Zhou}, {and}
  \bibinfo{person}{Junyang Lin}.} \bibinfo{year}{2024}\natexlab{a}.
\newblock \showarticletitle{Qwen2-VL: Enhancing Vision-Language Model's
  Perception of the World at Any Resolution}.
\newblock \bibinfo{journal}{\emph{arXiv preprint arXiv:2409.12191}}
  (\bibinfo{year}{2024}).
\newblock


\bibitem[Wang et~al\mbox{.}(2023)]%
        {wang2023math}
\bibfield{author}{\bibinfo{person}{Peiyi Wang}, \bibinfo{person}{Lei Li},
  \bibinfo{person}{Zhihong Shao}, \bibinfo{person}{RX Xu},
  \bibinfo{person}{Damai Dai}, \bibinfo{person}{Yifei Li},
  \bibinfo{person}{Deli Chen}, \bibinfo{person}{Yu Wu}, {and}
  \bibinfo{person}{Zhifang Sui}.} \bibinfo{year}{2023}\natexlab{}.
\newblock \showarticletitle{Math-shepherd: Verify and reinforce llms
  step-by-step without human annotations}.
\newblock \bibinfo{journal}{\emph{arXiv preprint arXiv:2312.08935}}
  (\bibinfo{year}{2023}).
\newblock


\bibitem[Wang et~al\mbox{.}(2024c)]%
        {wang2024math}
\bibfield{author}{\bibinfo{person}{Peiyi Wang}, \bibinfo{person}{Lei Li},
  \bibinfo{person}{Zhihong Shao}, \bibinfo{person}{Runxin Xu},
  \bibinfo{person}{Damai Dai}, \bibinfo{person}{Yifei Li},
  \bibinfo{person}{Deli Chen}, \bibinfo{person}{Yu Wu}, {and}
  \bibinfo{person}{Zhifang Sui}.} \bibinfo{year}{2024}\natexlab{c}.
\newblock \showarticletitle{Math-Shepherd: Verify and Reinforce LLMs
  Step-by-step without Human Annotations}. In
  \bibinfo{booktitle}{\emph{Proceedings of the 62nd Annual Meeting of the
  Association for Computational Linguistics}}. \bibinfo{pages}{9426--9439}.
\newblock


\bibitem[Wang et~al\mbox{.}(2024b)]%
        {wang2024enhancing}
\bibfield{author}{\bibinfo{person}{Weiyun Wang}, \bibinfo{person}{Zhe Chen},
  \bibinfo{person}{Wenhai Wang}, \bibinfo{person}{Yue Cao},
  \bibinfo{person}{Yangzhou Liu}, \bibinfo{person}{Zhangwei Gao},
  \bibinfo{person}{Jinguo Zhu}, \bibinfo{person}{Xizhou Zhu},
  \bibinfo{person}{Lewei Lu}, \bibinfo{person}{Yu Qiao}, {et~al\mbox{.}}}
  \bibinfo{year}{2024}\natexlab{b}.
\newblock \showarticletitle{Enhancing the reasoning ability of multimodal large
  language models via mixed preference optimization}.
\newblock \bibinfo{journal}{\emph{arXiv preprint arXiv:2411.10442}}
  (\bibinfo{year}{2024}).
\newblock


\bibitem[Wang et~al\mbox{.}(2025)]%
        {visualprm}
\bibfield{author}{\bibinfo{person}{Weiyun Wang}, \bibinfo{person}{Zhangwei
  Gao}, \bibinfo{person}{Lianjie Chen}, \bibinfo{person}{Zhe Chen},
  \bibinfo{person}{Jinguo Zhu}, \bibinfo{person}{Xiangyu Zhao},
  \bibinfo{person}{Yangzhou Liu}, \bibinfo{person}{Yue Cao},
  \bibinfo{person}{Shenglong Ye}, \bibinfo{person}{Xizhou Zhu},
  {et~al\mbox{.}}} \bibinfo{year}{2025}\natexlab{}.
\newblock \showarticletitle{VisualPRM: An Effective Process Reward Model for
  Multimodal Reasoning}.
\newblock \bibinfo{journal}{\emph{arXiv preprint arXiv:2503.10291}}
  (\bibinfo{year}{2025}).
\newblock


\bibitem[Wu et~al\mbox{.}(2025)]%
        {astar}
\bibfield{author}{\bibinfo{person}{Jinyang Wu}, \bibinfo{person}{Mingkuan
  Feng}, \bibinfo{person}{Shuai Zhang}, \bibinfo{person}{Ruihan Jin},
  \bibinfo{person}{Feihu Che}, \bibinfo{person}{Zengqi Wen}, {and}
  \bibinfo{person}{Jianhua Tao}.} \bibinfo{year}{2025}\natexlab{}.
\newblock \showarticletitle{Boosting Multimodal Reasoning with MCTS-Automated
  Structured Thinking}.
\newblock \bibinfo{journal}{\emph{arXiv preprint arXiv:2502.02339}}
  (\bibinfo{year}{2025}).
\newblock


\bibitem[Wu et~al\mbox{.}(2024)]%
        {wu2024deepseek}
\bibfield{author}{\bibinfo{person}{Zhiyu Wu}, \bibinfo{person}{Xiaokang Chen},
  \bibinfo{person}{Zizheng Pan}, \bibinfo{person}{Xingchao Liu},
  \bibinfo{person}{Wen Liu}, \bibinfo{person}{Damai Dai},
  \bibinfo{person}{Huazuo Gao}, \bibinfo{person}{Yiyang Ma},
  \bibinfo{person}{Chengyue Wu}, \bibinfo{person}{Bingxuan Wang},
  {et~al\mbox{.}}} \bibinfo{year}{2024}\natexlab{}.
\newblock \showarticletitle{Deepseek-vl2: Mixture-of-experts vision-language
  models for advanced multimodal understanding}.
\newblock \bibinfo{journal}{\emph{arXiv preprint arXiv:2412.10302}}
  (\bibinfo{year}{2024}).
\newblock


\bibitem[Xiao et~al\mbox{.}(2024)]%
        {xiao2024logicvista}
\bibfield{author}{\bibinfo{person}{Yijia Xiao}, \bibinfo{person}{Edward Sun},
  \bibinfo{person}{Tianyu Liu}, {and} \bibinfo{person}{Wei Wang}.}
  \bibinfo{year}{2024}\natexlab{}.
\newblock \showarticletitle{Logicvista: Multimodal llm logical reasoning
  benchmark in visual contexts}.
\newblock \bibinfo{journal}{\emph{arXiv preprint arXiv:2407.04973}}
  (\bibinfo{year}{2024}).
\newblock


\bibitem[Xu et~al\mbox{.}(2024)]%
        {llava-o1}
\bibfield{author}{\bibinfo{person}{Guowei Xu}, \bibinfo{person}{Peng Jin},
  \bibinfo{person}{Li Hao}, \bibinfo{person}{Yibing Song},
  \bibinfo{person}{Lichao Sun}, {and} \bibinfo{person}{Li Yuan}.}
  \bibinfo{year}{2024}\natexlab{}.
\newblock \showarticletitle{Llava-o1: Let vision language models reason
  step-by-step}.
\newblock \bibinfo{journal}{\emph{arXiv preprint arXiv:2411.10440}}
  (\bibinfo{year}{2024}).
\newblock


\bibitem[Yan et~al\mbox{.}(2022)]%
        {yan2022scale}
\bibfield{author}{\bibinfo{person}{Le Yan}, \bibinfo{person}{Zhen Qin},
  \bibinfo{person}{Xuanhui Wang}, \bibinfo{person}{Michael Bendersky}, {and}
  \bibinfo{person}{Marc Najork}.} \bibinfo{year}{2022}\natexlab{}.
\newblock \showarticletitle{Scale calibration of deep ranking models}. In
  \bibinfo{booktitle}{\emph{Proceedings of the 28th ACM SIGKDD Conference on
  Knowledge Discovery and Data Mining}}. \bibinfo{pages}{4300--4309}.
\newblock


\bibitem[Yang et~al\mbox{.}(2025)]%
        {yang2025r1}
\bibfield{author}{\bibinfo{person}{Yi Yang}, \bibinfo{person}{Xiaoxuan He},
  \bibinfo{person}{Hongkun Pan}, \bibinfo{person}{Xiyan Jiang},
  \bibinfo{person}{Yan Deng}, \bibinfo{person}{Xingtao Yang},
  \bibinfo{person}{Haoyu Lu}, \bibinfo{person}{Dacheng Yin},
  \bibinfo{person}{Fengyun Rao}, \bibinfo{person}{Minfeng Zhu},
  {et~al\mbox{.}}} \bibinfo{year}{2025}\natexlab{}.
\newblock \showarticletitle{R1-Onevision: Advancing Generalized Multimodal
  Reasoning through Cross-Modal Formalization}.
\newblock \bibinfo{journal}{\emph{arXiv preprint arXiv:2503.10615}}
  (\bibinfo{year}{2025}).
\newblock


\bibitem[Yao et~al\mbox{.}(2024a)]%
        {mulberry}
\bibfield{author}{\bibinfo{person}{Huanjin Yao}, \bibinfo{person}{Jiaxing
  Huang}, \bibinfo{person}{Wenhao Wu}, \bibinfo{person}{Jingyi Zhang},
  \bibinfo{person}{Yibo Wang}, \bibinfo{person}{Shunyu Liu},
  \bibinfo{person}{Yingjie Wang}, \bibinfo{person}{Yuxin Song},
  \bibinfo{person}{Haocheng Feng}, \bibinfo{person}{Li Shen}, {et~al\mbox{.}}}
  \bibinfo{year}{2024}\natexlab{a}.
\newblock \showarticletitle{Mulberry: Empowering mllm with o1-like reasoning
  and reflection via collective monte carlo tree search}.
\newblock \bibinfo{journal}{\emph{arXiv preprint arXiv:2412.18319}}
  (\bibinfo{year}{2024}).
\newblock


\bibitem[Yao et~al\mbox{.}(2024b)]%
        {yao2024mulberry}
\bibfield{author}{\bibinfo{person}{Huanjin Yao}, \bibinfo{person}{Jiaxing
  Huang}, \bibinfo{person}{Wenhao Wu}, \bibinfo{person}{Jingyi Zhang},
  \bibinfo{person}{Yibo Wang}, \bibinfo{person}{Shunyu Liu},
  \bibinfo{person}{Yingjie Wang}, \bibinfo{person}{Yuxin Song},
  \bibinfo{person}{Haocheng Feng}, \bibinfo{person}{Li Shen}, {et~al\mbox{.}}}
  \bibinfo{year}{2024}\natexlab{b}.
\newblock \showarticletitle{Mulberry: Empowering mllm with o1-like reasoning
  and reflection via collective monte carlo tree search}.
\newblock \bibinfo{journal}{\emph{arXiv preprint arXiv:2412.18319}}
  (\bibinfo{year}{2024}).
\newblock


\bibitem[Yao et~al\mbox{.}(2024c)]%
        {yao2024minicpm}
\bibfield{author}{\bibinfo{person}{Yuan Yao}, \bibinfo{person}{Tianyu Yu},
  \bibinfo{person}{Ao Zhang}, \bibinfo{person}{Chongyi Wang},
  \bibinfo{person}{Junbo Cui}, \bibinfo{person}{Hongji Zhu},
  \bibinfo{person}{Tianchi Cai}, \bibinfo{person}{Haoyu Li},
  \bibinfo{person}{Weilin Zhao}, \bibinfo{person}{Zhihui He}, {et~al\mbox{.}}}
  \bibinfo{year}{2024}\natexlab{c}.
\newblock \showarticletitle{Minicpm-v: A gpt-4v level mllm on your phone}.
\newblock \bibinfo{journal}{\emph{arXiv preprint arXiv:2408.01800}}
  (\bibinfo{year}{2024}).
\newblock


\bibitem[Zhang et~al\mbox{.}(2024d)]%
        {rest-mcts}
\bibfield{author}{\bibinfo{person}{Dan Zhang}, \bibinfo{person}{Sining
  Zhoubian}, \bibinfo{person}{Ziniu Hu}, \bibinfo{person}{Yisong Yue},
  \bibinfo{person}{Yuxiao Dong}, {and} \bibinfo{person}{Jie Tang}.}
  \bibinfo{year}{2024}\natexlab{d}.
\newblock \showarticletitle{Rest-mcts*: Llm self-training via process reward
  guided tree search}.
\newblock \bibinfo{journal}{\emph{Advances in Neural Information Processing
  Systems}}  \bibinfo{volume}{37} (\bibinfo{year}{2024}),
  \bibinfo{pages}{64735--64772}.
\newblock


\bibitem[Zhang et~al\mbox{.}(2024a)]%
        {zhang2024generative}
\bibfield{author}{\bibinfo{person}{Lunjun Zhang}, \bibinfo{person}{Arian
  Hosseini}, \bibinfo{person}{Hritik Bansal}, \bibinfo{person}{Mehran Kazemi},
  \bibinfo{person}{Aviral Kumar}, {and} \bibinfo{person}{Rishabh Agarwal}.}
  \bibinfo{year}{2024}\natexlab{a}.
\newblock \showarticletitle{Generative verifiers: Reward modeling as next-token
  prediction}.
\newblock \bibinfo{journal}{\emph{arXiv preprint arXiv:2408.15240}}
  (\bibinfo{year}{2024}).
\newblock


\bibitem[Zhang et~al\mbox{.}(2024b)]%
        {zhang2024mathverse}
\bibfield{author}{\bibinfo{person}{Renrui Zhang}, \bibinfo{person}{Dongzhi
  Jiang}, \bibinfo{person}{Yichi Zhang}, \bibinfo{person}{Haokun Lin},
  \bibinfo{person}{Ziyu Guo}, \bibinfo{person}{Pengshuo Qiu},
  \bibinfo{person}{Aojun Zhou}, \bibinfo{person}{Pan Lu},
  \bibinfo{person}{Kai-Wei Chang}, \bibinfo{person}{Yu Qiao}, {et~al\mbox{.}}}
  \bibinfo{year}{2024}\natexlab{b}.
\newblock \showarticletitle{Mathverse: Does your multi-modal llm truly see the
  diagrams in visual math problems?}. In \bibinfo{booktitle}{\emph{European
  Conference on Computer Vision}}. Springer, \bibinfo{pages}{169--186}.
\newblock


\bibitem[Zhang et~al\mbox{.}(2024c)]%
        {zhang2024improve}
\bibfield{author}{\bibinfo{person}{Ruohong Zhang}, \bibinfo{person}{Bowen
  Zhang}, \bibinfo{person}{Yanghao Li}, \bibinfo{person}{Haotian Zhang},
  \bibinfo{person}{Zhiqing Sun}, \bibinfo{person}{Zhe Gan},
  \bibinfo{person}{Yinfei Yang}, \bibinfo{person}{Ruoming Pang}, {and}
  \bibinfo{person}{Yiming Yang}.} \bibinfo{year}{2024}\natexlab{c}.
\newblock \showarticletitle{Improve vision language model chain-of-thought
  reasoning}.
\newblock \bibinfo{journal}{\emph{arXiv preprint arXiv:2410.16198}}
  (\bibinfo{year}{2024}).
\newblock


\bibitem[Zhang et~al\mbox{.}(2025)]%
        {qwenprm}
\bibfield{author}{\bibinfo{person}{Zhenru Zhang}, \bibinfo{person}{Chujie
  Zheng}, \bibinfo{person}{Yangzhen Wu}, \bibinfo{person}{Beichen Zhang},
  \bibinfo{person}{Runji Lin}, \bibinfo{person}{Bowen Yu},
  \bibinfo{person}{Dayiheng Liu}, \bibinfo{person}{Jingren Zhou}, {and}
  \bibinfo{person}{Junyang Lin}.} \bibinfo{year}{2025}\natexlab{}.
\newblock \showarticletitle{The lessons of developing process reward models in
  mathematical reasoning}.
\newblock \bibinfo{journal}{\emph{arXiv preprint arXiv:2501.07301}}
  (\bibinfo{year}{2025}).
\newblock


\bibitem[Zhou et~al\mbox{.}(2025)]%
        {zhou2025r1}
\bibfield{author}{\bibinfo{person}{Hengguang Zhou}, \bibinfo{person}{Xirui Li},
  \bibinfo{person}{Ruochen Wang}, \bibinfo{person}{Minhao Cheng},
  \bibinfo{person}{Tianyi Zhou}, {and} \bibinfo{person}{Cho-Jui Hsieh}.}
  \bibinfo{year}{2025}\natexlab{}.
\newblock \showarticletitle{R1-Zero's" Aha Moment" in Visual Reasoning on a 2B
  Non-SFT Model}.
\newblock \bibinfo{journal}{\emph{arXiv preprint arXiv:2503.05132}}
  (\bibinfo{year}{2025}).
\newblock


\bibitem[Zou et~al\mbox{.}(2024)]%
        {dynamath}
\bibfield{author}{\bibinfo{person}{Chengke Zou}, \bibinfo{person}{Xingang Guo},
  \bibinfo{person}{Rui Yang}, \bibinfo{person}{Junyu Zhang},
  \bibinfo{person}{Bin Hu}, {and} \bibinfo{person}{Huan Zhang}.}
  \bibinfo{year}{2024}\natexlab{}.
\newblock \bibinfo{title}{DynaMath: A Dynamic Visual Benchmark for Evaluating
  Mathematical Reasoning Robustness of Vision Language Models}.
\newblock
\newblock
\showeprint[arxiv]{2411.00836}~[cs.CV]
\urldef\tempurl%
\url{https://arxiv.org/abs/2411.00836}
\showURL{%
\tempurl}


\end{thebibliography}

\end{document}